\pdfoutput=1

\PassOptionsToPackage{prologue,dvipsnames}{xcolor}
\documentclass[sigconf]{acmart}

\AtBeginDocument{%
  }

\setcopyright{acmcopyright}
\copyrightyear{2018}
\acmYear{2018}
\acmDOI{XXXXXXX.XXXXXXX}

\acmConference[ICSE 2024]{46th International Conference on Software Engineering}{April 2024}{Lisbon, Portugal}

\acmPrice{15.00}
\acmISBN{978-1-4503-XXXX-X/18/06}

\usepackage[final]{changes}

\usepackage{xcolor}
\usepackage{booktabs}
\usepackage{multirow}
\usepackage{makecell}
\usepackage{fbox} 
\usepackage{fancybox} 
\usepackage{colortbl}
\usepackage{graphicx}
\usepackage{enumitem}
\usepackage{xspace}
\usepackage{tcolorbox}

\begin{document}

\title{When Contracts Meets Crypto: Exploring Developers' Struggles with Ethereum Cryptographic APIs}

\author{Jiashuo Zhang}
\affiliation{%
  \institution{School of Computer Science \\Peking University}
  \city{Beijing}
  \country{China}
}
\email{zhangjiashuo@pku.edu.cn}

\author{Jiachi Chen}
\authornote{corresponding author}
\affiliation{%
  \institution{Sun Yat-sen University}
  \city{Zhuhai}
  \country{China}
}
\email{chenjch86@mail.sysu.edu.cn}

\author{Zhiyuan Wan}
\affiliation{%
  \institution{Zhejiang University}
  \city{Hangzhou}
  \country{China}
}
\email{wanzhiyuan@zju.edu.cn}

\author{Ting Chen}
\affiliation{
  \institution{University of Electronic Science and Technology of China}
  \city{Chengdu}
  \country{China}
}
\email{brokendragon@uestc.edu.cn}

\author{Jianbo Gao}
\affiliation{%
  \institution{School of Computer Science \\Peking University}
  \city{Beijing}
  \country{China}
}
\email{gaojianbo@pku.edu.cn}

\author{Zhong Chen}
\affiliation{%
  \institution{School of Computer Science \\Peking University}
  \city{Beijing}
  \country{China}
}
\email{zhongchen@pku.edu.cn}

\renewcommand{\shortauthors}{Zhang et al.}

\newcommand{\etal}{{\emph{et al.}}\xspace}
\newcommand{\eg}{{\emph{e.g.}}\xspace}
\newcommand{\ie}{{\emph{i.e.}}\xspace}
\newcommand{\etc}{{\emph{etc.}}\xspace}

\newcommand{\SHA}{$\mathtt{KECCAK256}$\xspace}
\newcommand{\KECCAK}{$\mathtt{KECCAK256}$\xspace}
\newcommand{\Ecrecover}{$\mathtt{ECRECOVER}$\xspace}
\newcommand{\SHATWO}{$\mathtt{SHA256}$\xspace}
\newcommand{\RIPEMD}{$\mathtt{RIPEMD160}$\xspace}
\newcommand{\MODEXP}{$\mathtt{MODEXP}$\xspace}
\newcommand{\ECADD}{$\mathtt{ECADD}$\xspace}
\newcommand{\ECMUL}{$\mathtt{ECMUL}$\xspace}
\newcommand{\ECPAIRING}{$\mathtt{ECPAIRING}$\xspace}
\newcommand{\BLAKE}{$\mathtt{BLAKE2F}$\xspace}

\newcommand{\myparagraph}[1]{{ \textbf{#1.}}\quad }

\newcommand{\obs}[2]{  \begin{tcolorbox}[colframe=black, colback=gray!10, boxrule=1pt, left=1pt,right=1pt,top=3pt,bottom=3pt]  \setlength{\parindent}{0em} \textit{\textbf{Observation #1: }#2} \end{tcolorbox}  }

\begin{abstract}
To empower smart contracts with the promising capabilities of cryptography, Ethereum officially introduced a set of cryptographic APIs that facilitate basic cryptographic operations within smart contracts, such as elliptic curve operations.
However, since developers are not necessarily cryptography experts, requiring them to directly interact with these basic APIs has caused real-world security issues and potential usability challenges.
To guide future research and solutions to these challenges, we conduct the first empirical study on Ethereum cryptographic practices.
Through the analysis of 91,484,856 Ethereum transactions, 500 crypto-related contracts, and 483 StackExchange posts, we provide the first in-depth look at cryptographic tasks developers need to accomplish and identify five categories of obstacles they encounter.
Furthermore, we conduct an online survey with 78 smart contract practitioners to explore their perspectives on these obstacles and elicit the underlying reasons.
We find that more than half of practitioners face more challenges in cryptographic tasks compared to general business logic in smart contracts.
Their feedback highlights the gap between low-level cryptographic APIs and high-level tasks they need to accomplish, emphasizing the need for improved cryptographic APIs, task-based templates, and effective assistance tools.
Based on these findings, we provide practical implications for further improvements and outline future research directions.

\end{abstract}


\maketitle

\section{Introduction}
Cryptographic techniques, leveraging their capabilities to ensure the security of data, computation, and communication, have brought a wide range of innovations to smart contracts.
By employing advanced cryptographic tools, smart contracts have effectively promoted on-chain identity authentication~\cite{eip-712,eip-2612}, private computation~\cite{belles2022circom, steffen2019zkay, steffen2022zeestar}, and numerous other promising functionalities~\cite{ wang2023zero, zkroolup, zhang2021boros}. These capabilities significantly enhance the flexibility and operational scope of on-chain applications.

To enable on-chain cryptographic practices, Ethereum officially introduces a set of system-level cryptographic APIs and allows developers to implement various cryptographic tasks based on them.
For example, the verification of ECDSA signatures~\cite{johnson2001elliptic} can be implemented by composing the \KECCAK and \Ecrecover APIs~\cite{wood2014ethereum} in sequence.
These crypto APIs effectively reduce the gas cost associated with complex cryptographic operations and attract emerging applications from the industry.

However, while the advent of Ethereum crypto APIs theoretically facilitates on-chain cryptographic practices, 
the practical challenges related to their utilization and integration may hinder developers from accomplishing their cryptographic tasks.
Since smart contract developers may not necessarily be cryptographic experts, expecting them to directly interact with fundamental cryptographic operations such as elliptic curve addition could pose usability challenges and potential misuse risks.
For example, a security team reported that 52 smart contracts suffer signature replay attacks due to their vulnerable signature verification implementations~\cite{defcon}.
To prevent such difficulties from constraining the broader on-chain application of cryptographic tools,
it is essential to understand the obstacles developers encounter and provide better support for them.

\added{While usability of traditional crypto libraries has been well-studied~\cite{nadi2016jumping,patnaik2019usability,mindermann2018usable}, the real-world usage of Ethereum crypto APIs remains unexplored.
This paper aims to characterize obstacles in using Ethereum crypto APIs, elicit underlying reasons behind them, and derive potential solutions to mitigate them.}
Specifically, we focus on the following research questions.

\vspace{0.3em}
\noindent{\textbf{\textit{RQ1. What are the common cryptographic practices in smart contracts?}}}
\vspace{0.3em}

Understanding the tasks developers need to perform is the first step to understand the obstacles they face~\cite{robillard2009makes,nadi2016jumping}.
To characterize cryptographic practices in smart contracts, we conducted a study on 91,484,856 Ethereum transactions and 500 crypto-related smart contracts.
Our findings reveal the prevalence, classification, and distribution of common cryptographic tasks in Ethereum smart contracts, including signature~\cite{nist1992digital}, commitment~\cite{damgaard1998commitment}, message digest~\cite{preneel1994cryptographic}, random number generator~\cite{random-number}, proof of work~\cite{proof-of-work}, and zero-knowledge proof~\cite{thaler2022proofs}.
Such information can guide solution designs and align community support with real-world demands.

\vspace{0.3em}
\noindent\textbf{\textit{RQ2. What obstacles, if any, do developers face when accomplishing on-chain cryptographic tasks?}}
\vspace{0.3em}

Existing research highlighted that developers struggle with cryptographic tasks in traditional applications~\cite{green2016developers,nadi2016jumping,mindermann2018usable}, but the challenges specific to smart contract cryptographic practices remain unknown.
Through the analysis of 483 StackExchange posts, we confirmed that smart contract developers face obstacles in cryptographic practices. We further classified these obstacles into five categories, i.e., \textit{Knowledge, Roadmap Identification, Template Usage, API Usage}, and \textit{Security}.
These obstacles underline the challenges in understanding, implementing, and securing on-chain cryptographic practices, and highlight the need to provide better support for developers.

\vspace{0.3em}
\noindent\textbf{\textit{RQ3. How do real-world practitioners perceive these obstacles and what improvements do they desire?}}
\vspace{0.3em}

While our analysis of RQ2 confirmed the presence of challenges in smart contract cryptographic practices, the root causes of these obstacles and possible solutions to mitigate them remain unexplored.
To understand obstacles and solutions from the practitioners' perspective, 
we conducted an online survey involving 78 smart contract practitioners.
The results suggest that the current crypto APIs might be too low-level for developers.
57.8\% of participants face obstacles in identifying detailed implementation steps for specific cryptographic tasks, \eg, deciding which crypto API should be used, before they actually operate the underlying APIs.
The practitioners' feedback highlights the need for improved crypto APIs, task-based templates, effective testing/audit tools, and easy-to-understand documentation.

In the following, we summarize our main contributions:
\vspace{-0.5em}
\begin{itemize}
\item %
We categorize common cryptographic tasks in smart contracts and explore their prevalence and distribution. The results provide the first close look at on-chain cryptographic practices and can derive guidance for further studies.
\item We reveal the obstacles faced by developers in accomplishing on-chain cryptographic tasks and investigate them from the practitioners' perspective.
Our findings shed light on the underlying causes of these obstacles and provide insights for future tools and solution designs.
\item We publish our datasets and results at 
\added{https://zenodo.org/ records/10074040}
to facilitate further studies.
\end{itemize}

\section{Background}

\subsection{Ethereum Virtual Machine (EVM)}
The Ethereum Virtual Machine (EVM) serves as the runtime environment to execute Ethereum smart contracts~\cite{wood2014ethereum}.
During the execution of a contract, EVM executes a number of specific operations known as opcodes, modifies its state (including the stack, memory, and storage), and records the execution results onto the blockchain.

\myparagraph{Opcode}
Specifically, each opcode refers to a specific operation that transitions the EVM from the current state to the next.
These operations could involve writing data to the blockchain, reading the blockchain state, calling another contract, and other operations.
For example, by using the $\mathtt{STATICCALL}$/$\mathtt{
	CALL}$/ $\mathtt{
	CALLCODE}$/$\mathtt{
	DELEGATECALL}$ opcode, a smart contract can interact with other contracts and execute their codes.
Using the $\mathtt{SLOAD}$/$\mathtt{SSTORE}$ opcode, the smart contract can read the data from the EVM storage to the stack or store the data from the stack to the storage.
The detailed definition of each opcode is available in the Ethereum yellow paper~\cite{wood2014ethereum}.

\myparagraph{Precompiled Contracts}
Except for these opcodes, Ethereum also introduces a set of precompiled contracts~\cite{wood2014ethereum} to support other advanced functionalities, such as cryptographic operations in smart contracts.
Different from user-defined smart contracts implemented in Solidity, precompiled contracts are \emph{low-level extensions} of EVM, implemented in the same languages as EVM (\eg, Go or C++).
In this case, the computation cost of the precompiled contracts can be highly optimized, thus making complex cryptographic operations feasible in smart contracts.
Any contract can use the $\mathtt{STATICCALL}$/$\mathtt{
	CALL}$/$\mathtt{
	CALLCODE}$/$\mathtt{
	DELEGATECALL}$ opcode to call precompiled contracts and execute their functionalities.

\subsection{Crypto APIs in EVM}
To reduce the cost of cryptographic operations, 
Ethereum provides a set of system-level crypto APIs in EVM.
Specifically, there are one EVM opcode (\ie, \KECCAK) and eight precompiled contracts~\cite{wood2014ethereum} (\ie, \Ecrecover,  \SHATWO, \RIPEMD, \MODEXP , \ECADD, \ECMUL, \ECPAIRING,  \BLAKE), providing cryptographic functionalities to smart contracts.

According to the Ethereum yellow paper~\cite{wood2014ethereum} and Ethereum improvement proposals (EIP-152~\cite{eip-152}, EIP-196~\cite{eip-196}, EIP-197~\cite{eip-197}, EIP-198~\cite{eip-198}), current APIs mainly provide basic operations for hash, signature, and zero-knowledge proof (ZKP) tasks.
Specifically, \KECCAK, \SHATWO, \RIPEMD, \BLAKE provide four common hash functions~\cite{wood2014ethereum,eip-152}, \Ecrecover and \MODEXP enable ECDSA and RSA signature verification~\cite{wood2014ethereum,eip-198}, and \ECADD, \ECMUL, \ECPAIRING are officially stated as ``ZKP-related precompiled contracts''~\cite{eip-196,eip-197}.

\label{sec:cryptoapis}

\myparagraph{KECCAK256, SHA256, RIPEMD160, BLAKE2F}
These four APIs enable four different hash functions in smart contracts.
Specifically, 
the \KECCAK opcode was introduced since Frontier (2015.07)~\cite{frontier}, computing KECCAK256~\cite{keccak} hash functions in smart contracts.
The \SHATWO and \RIPEMD precompiled contracts were introduced since Frontier~\cite{frontier}, implementing the standard SHA256~\cite{penard2008secure} and RIPEMD160~\cite{dobbertin1996ripemd} hash functions, respectively.
The \BLAKE was introduced since the Istanbul hardfork (2019.11)~\cite{istanbul}, to support interoperability between EVM and Zcash~\cite{hopwood2016zcash} and also introduce an alternative SHA3-finalist hash function~\cite{eip-152,sha3contest}.

\
\myparagraph{ECRECOVER, MODEXP}
These two precompiled contracts enable efficient signature verification in smart contracts.
Specifically, \Ecrecover provides the public key recovery functionality for ECDSA signatures~\cite{johnson2001elliptic} based on the SECP-256k1 elliptic curve~\cite{secp256k1}. It was introduced since Frontier ~\cite{frontier} to enable on-chain ECDSA signature verification.
\MODEXP was introduced since the Byzantine hardfork (2017.10)~\cite{byzantium}.
It provides big integer modular exponentiation operations to support efficient RSA signature verification~\cite{rivest1978method}, and other number-theory-based cryptography~\cite{eip-198}.

\myparagraph{ECADD, ECMUL, ECPAIRING}
\ECADD, \ECMUL, and \ECPAIRING were introduced since the Byzantine hardfork~\cite{byzantium}. They provide addition, scalar multiplication, and pairing operations on the \emph{alt\_bn128} elliptic curve respectively to support pairing-based zero-knowledge proof systems such as Groth16~\cite{groth2016size} and Plonk~\cite{gabizon2019plonk}.

\subsection{Card Sorting Approach}
\added{To analyze the contract source codes and StackExchange posts, we employed card sorting, a common approach for organizing information into logical groups~\cite{wood2008card}.}
Initially, users create a card for each item, including any relevant information for classification purposes. Then they assign each card to an appropriate category in a bottom-up manner.
There are three types of card sorting based on whether the categories are predefined: open card sorting, closed card sorting, and hybrid card sorting. In closed card sorting, all categories are predefined, while in open card sorting, users define the categories themselves. The hybrid card sorting is a compromise between open and closed card sorting: it has predefined categories while also allowing users to create new ones.

\section{Answer to RQ1: Cryptographic practices in the wild}
\label{sec:RQ1}

Exploring common practices that developers need to perform helps to understand the challenges they encounter and provide specific support~\cite{robillard2009makes, nadi2016jumping}.
In this section, we conducted the first study of on-chain cryptographic practices.
First, by replaying 91,484,856 Ethereum historical transactions, we investigated the real-world usage of Ethereum crypto APIs and explored the prevalence of each crypto API.
Second, by manually analyzing 500 crypto-related smart contracts, we categorized common cryptographic tasks developers need to implement and evaluated their distribution.

\subsection{Data Collection and Processing}

\myparagraph{Transaction Execution Data}
To evaluate the prevalence of each crypto API, we collected the execution data of Ethereum transactions and analyzed crypto API calls within them.
We randomly sampled 91,484,856 transactions from all 1,928,853,563 transactions on Ethereum from 1 to 17,000,000 blocks (2015.07 to 2023.04), representing the entire set with a confidence level of 95\% and a confidence interval of 0.01\%.
We used the \emph{debug\_traceTransaction} API of Go-Ethereum~\cite{debugapi} to replay all sampled transactions and get their execution traces,
including the executed opcodes and the internal states (\emph{i.e.}, stack, memory, and storage) of EVM.
Then, we analyzed the execution traces to identify the calls to crypto APIs.

Specifically, we used different methods to identify calls to the crypto-related opcode and precompiled contracts.
For the eight crypto-related precompiled contracts, we analyzed the destination address of all contract call opcodes ($\mathtt{STATICCALL}$, $\mathtt{
	CALL}$, $\mathtt{
	CALLCODE}$, and $\mathtt{
	DELEGATECALL}$) to determine whether they call these contracts.
For example, the $\mathtt{
CALL}$ opcode takes the top seven elements of the EVM stack as input, where the second element is the destination contract that it intends to call.
For the \KECCAK opcode, we only filtered \KECCAK cryptographic operations.
While \KECCAK is natively used in Solidity to maintain the mapping and dynamic array data types and compute the topics for on-chain events~\cite{wood2014ethereum}, we only focused on user-initiated cryptographic operations, instead of these native operations initiated by the Solidity compiler. Specifically, we filtered out native operations based on their operation characteristics.
For example, for a mapping variable $data$ with type $mapping(unit\Rightarrow address)$, access to $data[k]$ will be interpreted as the access to the storage slot located at $\mathtt{KECCAK256}(k\cdot p)$, where $p$ is the storage placeholder to uniquely identify the mapping and $\cdot$ is bytes concatenation. Therefore, if the output of \KECCAK is used as the key for storage operations ($\mathtt{SLOAD}$ and $\mathtt{SSTORE}$), it is a native operation to access the mapping and should be excluded from further analysis.
The detailed patterns to identify all native operations are available in the online supplemental material~\cite{supplement}.

As a result, we successfully identified 12,608,469 (13.8\%) crypto-related transactions out of 91,484,856 transactions.
The million-level transaction volume highlights the prevalence of cryptographic practices in smart contracts and the need for in-depth studies.

\myparagraph{Contract Source Codes}
To investigate the classification and distribution of cryptographic tasks in smart contracts, we collected source codes of crypto-related contracts for further analysis. 
First, we analyzed all contract calls in the transaction execution traces to extract contracts that initiated crypto API calls.
Then, we queried Etherscan~\cite{etherscan} to extract publicly available source codes of these contracts.
During the execution of 91,484,856 transactions, we identified 87,538 crypto-related contracts, 19,243 of which have publicly available source codes on Etherscan. \added{We did not deduplicate these contracts to faithfully characterize real-world distributions.}

\subsection{Results}
\subsubsection{The usage of crypto APIs}
\label{sec:prevalence}

We conducted a quantitative analysis of the proportion of transactions using each crypto API.
Table~\ref{tab:apis} shows the proportion of transactions using each crypto API.
In particular, \KECCAK and \Ecrecover, being used in 13.1\% and 4.9\% transactions respectively, are the two most prevalent crypto APIs.
It demonstrates that basic cryptographic tasks, such as KECCAK256 hash functions and ECDSA signatures, have already been extensively used in real-world on-chain applications.

\obs{1}{
The two most commonly used Ethereum crypto APIs are \KECCAK and \Ecrecover, utilized by 13.03\% and 4.96\% of transactions, respectively.
}

Compared to \KECCAK and \Ecrecover, the usage of other crypto APIs is relatively low (used in fewer than 1\% transactions), indicating that cryptographic tasks other than hash and signatures are less prevalent.
For example, while ZKP applications supported by \ECADD, \ECMUL, and \ECPAIRING have shown promising potential to enhance the scalability and privacy of smart contracts~\cite{zkroolup,belles2022circom, eberhardt2018zokrates}, their on-chain applications are still in the early stages.

\begin{table}[t]
	\centering
	\caption{Proportion of transactions using each crypto API}
 \vspace{-1em}

	\label{tab:apis}
	\begin{tabular}{llc}
		\toprule
		\textbf{Crypto API}& \textbf{API Interface} & \textbf{Proportion (\%)}\\ \midrule
  		\KECCAK           &  EVM opcode (0x20)  & 13.037                    \\ 
		\rowcolor{gray!30} 	\Ecrecover &  precompiled contract (0x1)& 4.958                  \\
    		 \SHATWO          &  precompiled contract (0x2)        & 0.554                   \\
\rowcolor{gray!30} \RIPEMD &  precompiled contract (0x3)     & 0.034                   \\
		  \MODEXP   & precompiled contract (0x5)    & 0.034                    \\
   	\rowcolor{gray!30}	\ECADD          &   precompiled contract (0x6)   & 0.022                 \\

		 \ECMUL    &  precompiled contract (0x7)          & 0.022                   \\
	\rowcolor{gray!30}		\ECPAIRING     & precompiled contract (0x8)                      & 0.022  \\ 
   \BLAKE         & precompiled contract (0x9)                  & 0.000   \\ 
		 \bottomrule
	\end{tabular}
 \vspace{-1em}
\end{table}

In particular, we found that the \BLAKE precompiled contract has rarely been used.
Since being introduced in the Istanbul hardfork (2019.11)~\cite{istanbul-upgrade},
it has only been used by four unique contracts in 52 transactions.
We analyzed these four contracts to understand the underlying reasons. Surprisingly,
\added{while EIP-152~\cite{eip-152} stated the envisioned usage of \BLAKE as "allowing interoperability between EVM and Zcash", and "introducing more hash primitives", these four contracts only uses the latter functionality. It suggests that \BLAKE, as well as on-chain SPV client~\cite{eip-152,nakamoto2008bitcoin} envisioned by it, might not be essential for real-world interoperability requirements. Moreover, compared to implementing $\mathtt{BLAKE}$-family hash functions based on \BLAKE, developers tend to adopt ready-to-use hash functions like \KECCAK.
The failure of the envisioned use cases and usability issues might cause the infrequent usage of \BLAKE. Notably, there is even a draft-status EIP (EIP-7266)~\cite{eip-7266} proposing its removal.}

\obs{2}{Since being introduced, the \BLAKE API has only been used by four smart contracts. It might suggest an information asymmetry between the API designer and real-world demands.}

\subsubsection{Common Cryptographic Tasks}
\label{sec:common-task}
\
To analyze the classification and distribution of common cryptographic tasks, we conducted a manual analysis of the source codes of crypto-related contracts and identified the tasks they implemented.
Due to the large dataset and the time-consuming analysis process, we sampled 500 out of 19,243 crypto-related contracts to make the manual analysis feasible.
On average, these contracts held 118.9 ETH and were called by 24,525 external transactions.

During manual analysis, two authors used the open card sorting approach to create categories for cryptographic tasks. %
Specifically, they followed the detailed workflow adopted in several previous studies~\cite{chen2020defining, yang2023definition}. First, they randomly examined 40\% contracts, identified the tasks implemented, and grouped them into relevant categories. If refining the categories is necessary, they went back to the beginning and reassigned these contracts to the new categories.
After that, they independently classified the remaining 60\% contracts.
The agreement of their results, measured by the kappa score~\cite{cohen1960coefficient}, is 0.86, demonstrating a high agreement.
\added{When a disagreement arises, they recheck their results and discuss to reach agreement on which result is more appropriate.}
As a result, we identified the following common cryptographic tasks.

\begin{table}[t]
	\centering
	\caption{Common cryptographic tasks in 500 Ethereum smart contracts using crypto APIs}
 \vspace{-1em}

	\label{tab:tasks}
	\begin{tabular}{lc}
		\toprule
		\textbf{Tasks}& \textbf{\# Contracts}\\ \midrule
  		Digital Signature                    & 197                    \\ 
		\rowcolor{gray!30} 	Commitment & 130                   \\
    		 Message Digest                   & 87                   \\

		 \rowcolor{gray!30}Random Number Generator    & 74                      \\
   		 Zero-knowledge Proof                  & 9                    \\

	\rowcolor{gray!30} Proof of Work                  & 8                  \\
			Other                   & 58                    \\ 
		 \bottomrule
	\end{tabular}

\end{table}

\myparagraph{Hash-based Tasks}
While Ethereum crypto APIs only provides basic hash functions, the developers create even more diverse cryptographic tasks based on them.
In particular, we identified four types of hash-based cryptographic tasks, namely \emph{commitment}, \emph{message digest}, \emph{random number generator}, and \emph{proof of work}.
They are implemented based on four hash crypto APIs, \emph{i.e.}, \KECCAK, \SHATWO, \RIPEMD, and \BLAKE.

\begin{itemize}[leftmargin=*]
\item \emph{commitment}. We identified two types of hash-based \emph{commitment} schemes, \emph{i.e.,} \emph{vector commitment} (121, 24.2\%) and \emph{single-value commitment} (9, 1.8\%).
The former refers to Merkle Tree (and its variants) which
allows committing to a vector of values and revealing subgroup values at specific positions later. It is often used for on-chain access control: the authorizer commits to the Merkle root, the entities who know a valid leaf and the corresponding Merkle proof can do sensitive operations, such as receiving airdrop tokens.
The latter refers to a heuristic approach that computes the hash of the value as the commitment and reveals the value itself later.
\item \emph{message digest}. \emph{message digest} is the most direct usage of hash functions, mapping an arbitrary-length message to a fixed-length ``digest''.
It has been implemented in 87 (17.4\%) contracts to compute constant-length storage/query indexes for arbitrary-length variables, \eg, the unique identifier of a token trading order.
\item \emph{random number generator}. 74 (14.8\%) contracts use hash functions as random number generators for scenarios such as NFT, gaming and gambling. They heuristically take attributes such as the block timestamp as seeds and hash the seeds to \emph{deterministically} generate random numbers. However, since the seed is predictable to front-running miners, the random numbers generated by such schemes might be weak and insecure~\cite{swc-120}.
\item \emph{proof of work (PoW)}. 8 (1.6\%) contracts introduced hash-based PoW schemes~\cite{proof-of-work}. Users need to solve hash-based puzzles to prove the expended computing effort~\cite{proof-of-work} before they perform certain operations.
For example, to prevent sybil attacks, token contracts might require a PoW before users mint a token.
\end{itemize}

\

\myparagraph{Digital Signature}
Signature verification tasks enable contracts to authenticate messages from off-chain entities and enforce access control policies for on-chain operations.
We found signature verification tasks in 197 (39.4\%) contracts, including ECDSA and RSA signatures. 
They usually follow the hash-then-sign paradigm: first, use hash crypto APIs to compute the hash of the signed message, then verify whether the signature is valid for the hash.

\myparagraph{Zero-knowledge Proof}
Zero-knowledge proof allows smart contracts to verify the results of off-chain computations without accessing the off-chain private inputs or re-executing the entire computation.
They have attracted emerging attention~\cite{eberhardt2018zokrates}~\cite{belles2022circom} to improve the scalability and privacy of on-chain computations.
We identified 9 (1.8\%) contracts that implement ZKP schemes, involving \ECADD, \ECMUL, \ECPAIRING, and \MODEXP crypto APIs.

\myparagraph{Other}
In addition, we
included an ``Other'' category to group infrequent (less than 1\%)
tasks we identified, such as the Chainlink VRF~\cite{kaleem2021demystifying}.
We also found two \emph{Solidity-specific tasks} in 55 (11.0\%) contracts, \emph{i.e.}, using hash APIs to compute function selectors (28, 5.6\%) and the addresses of contracts created by the $\mathtt{CREATE2}$ opcode~\cite{eip-1014} (27, 5.4\%).
Since they are \emph{solidity-specific} rather than \emph{common} cryptographic tasks, we also classified them into the Other category.

\obs{3}{There exists diverse cryptographic tasks in Ethereum smart contracts, including digital signature, commitment, message digest, random number generator, proof of work, and zero-knowledge proof.}

\section{Answer to RQ2: Identifying Obstacles from Q\&A Posts}
\label{sec:RQ2-StackExchange-Posts}

To check whether developers encounter obstacles and identify potential obstacles they face, we conducted an analysis of the posts from two influential Q\&A websites, \ie, Ethereum StackExchange~\cite{ethereumstackexchange} and StackOverflow~\cite{stackoverflow}.

\subsection{Data Collection and Processing}
\label{sec:s3-data-collection}
During data collection, we collected questions from Ethereum StackExchange and StackOverflow.
For Ethereum StackExchange, we manually filtered out 15 crypto-related tags (available in our supplement materials) from a total of 525 tags and included questions with these specific tags. For StackOverflow, we focused on questions tagged ``Solidity'' or ``smartcontracts''.
Overall, we collected 27,264 questions, including 19,717 Ethereum StackExchange questions and 7,547 StackOverflow questions. \added{All these questions are collected via the StackExchange Search API~\cite{searchapi}.} As both websites are affiliated with the StackExchange site network~\cite{stackexchange}, we collectively refer to the posts on these two websites as "StackExchange posts".

Then, we conducted a keyword-based filtering to extract the questions related to on-chain cryptographic practices.
To preserve the potential correlation between the obstacles and the crypto tasks developers try to perform, we separately collected questions for different categories of tasks.
Specifically, hash-based tasks are grouped into the same category since they involve the same APIs and similar concepts. Signatures and zero-knowledge proofs are classified as the other two categories.
We used the names of tasks and APIs such as "signature" and "ecrecover" as keywords to filter out questions that contained at least one of these keywords in their title or content.
Finally, we obtained 656 Hash-related questions, 415 Signature-related questions, and 44 ZKP-related questions.
To provide a 95\% confidence level and a 5\% confidence interval,
we randomly sampled 243, 200, and 40 questions for Hash, Signature, and ZKP-related questions, respectively.

\label{sec:s3-data-analysis}

Due to the exploratory nature of our study, we did not have any pre-defined categories for obstacles. Instead, we followed the same open card sorting procedure as Section~\ref{sec:common-task} to categorize these questions.
Specifically, two authors first randomly chose 40\% cards, identified the main obstacle the poster was likely facing, and classified each post according to the identified obstacle. 
They also introduced an N/A category to include cards that are not related to on-chain cryptographic practices.
Then, they independently categorized the remaining cards and discussed to resolve any differences in their results.
The agreement of the two authors, measured by the kappa score, is 0.83, demonstrating a high agreement~\cite{cohen1960coefficient}.

During data analysis, we noticed a relatively high proportion of N/A questions for Hash and Signature categories.
This is because keywords such as ``hash'' and ``signature'' have multiple meanings, including those unrelated to on-chain cryptographic practices.
For example, ``hash'' might refer to the transaction hashes, and ``signature'' might refer to function signatures defining the input and output of functions.  However, these concepts were irrelevant to our study. We excluded the N/A category from the subsequent analysis.

\begin{table}[]
	\centering
	\caption{Identified obstacles in StackExchange posts }
 \vspace{-1em}
	\label{tab:posts}
	\begin{tabular}{lccc}
		\toprule
		\textbf{Obstacles}& \textbf{Hash} & \textbf{Signature} & \textbf{ZKP}  \\ \midrule
		Knowledge                   & 11    & 13        & 7                  \\
		\rowcolor{gray!30} 	\makecell[l]{Roadmap Identification} & 20    & 32        & 17                   \\
		Template Usage              & 3    & 39        & 5                    \\
		\rowcolor{gray!30} API Usage                   & 51   & 30        & 2                    \\
		Security                    & 10    & 7         & 3                    \\ 
		\rowcolor{gray!30} 	N/A                    & 148    & 79        & 6                    \\ \midrule
		\# Analyzed Posts                    & 243   & 200        & 40                   \\ \bottomrule
	\end{tabular}
  \vspace{-1em}
\end{table}

\subsection{Results}
\label{sec:post-obstacles}
We identified five categories of obstacles from StackExchange posts and listed the number of posts facing each obstacle in Table~\ref{tab:posts}.
\added{The categorization of obstacles is in line with previous studies~\cite{nadi2016jumping,patnaik2019usability}, with unique categories introduced by the specific context of Ethereum.}
Specifically, the meaning of these obstacles is as follows.
\begin{itemize}[leftmargin=*]
    \item Knowledge: Lack of knowledge of cryptography or blockchain.
    \item Roadmap Identification: Difficulties in identifying the detailed steps to implement a specific task.
    \item Template Usage: Difficulties in using third-party templates. 
    \item API Usage: Difficulties in operating Ethereum crypto APIs.
    \item Security: Security concerns about the implementations.
\end{itemize}

We found that a portion of questions (11.6\%, 10.7\%, and 20.6\%, of Hash, Signature, and ZKP questions, respectively) were caused by insufficient knowledge of cryptography or blockchain.
It prevents posters from identifying the appropriate cryptographic tasks they need and assessing the feasibility of implementing these tasks on-chain.
For example, post 56124326~\cite{q56124326q} asked how to decode a SHA256 hash and retrieve its pre-image in Solidity. It reflects how a lack of understanding of hash led to the selection of incorrect tasks.

\obs{4}{
11.6\% Hash-related questions, 10.7\% Signature-related questions, and 20.6\% ZKP-related questions are caused by insufficient knowledge in cryptography or blockchain. 
}

Despite knowing the cryptographic tasks they need, developers might still encounter obstacles
when attempting to implement them.
Specifically, we identified three types of obstacles related to implementation issues, namely \textit{Roadmap Identification}, \textit{Template Usage}, and \textit{API Usage} in Table~\ref{tab:posts}. They are the main obstacles faced by 77.9\% Hash-related questions, 83.5\% Signature-related questions, and 70.6\% ZKP-related questions.

The \textit{Roadmap Identification} obstacle characterizes difficulties in identifying detailed steps to implement a specific task, \eg, identifying the crypto API or templates to use.
For example, post 73744~\cite{73744} asked for code examples and tutorials for achieving RSA signature verification within Solidity.
After reviewing this question, we found that the poster is well aware that RSA signature can be implemented on-chain, but does not know the specific implementation steps.

Even after identifying the detailed implementation steps, the crypto API complexity or unclear underlying implementation might still prevent developers from implementing them as planned.
Specifically, depending on whether these obstacles arise from reusing existing code templates or directly using the crypto APIs, we classified them into two categories, \ie, \textit{Template Usage} and \textit{API usage}.
\textit{Template Usage} obstacles include difficulties in understanding how third-party templates work and solving error messages/unexpected results when reusing them.
For example, in post 112807~\cite{112807}, the poster inquired about encountering unexpected results while using the OpenZepplin's ECDSA template~\cite{openzepplinECDSA}.
\textit{API usage} obstacles refer to problems in using Ethereum crypto APIs, \eg, what is the meaning of API parameters, and how to call the API in Solidity.
For example, in post 49261~\cite{49261} titled \textit{"How to use ecrecover() and what it is?"}, the poster asked what the parameters $v$, $r$, and $s$ of \Ecrecover mean and how to extract them from a signature.

We observed that the distribution of the three implementation-related obstacles differ across different cryptographic tasks.
For hash-related questions, the main obstacle is \emph{API usage}, accounting for 68.9\% of implementation issues. However, for signature and ZKP related questions, it only accounts for 29.7\% and 8.3\% of implementation issues, respectively.
A possible explanation is: for simple tasks like Hash, developers often have a clear understanding of the detailed implementation steps and do not need to use templates. Therefore, their main challenges lie in the direct usage of the APIs.
However, for tasks such as signature and ZKP that may involve compositions of different APIs, developers need more high-level guidance, such as identifying the correct sequence of API calls and understanding how existing templates work.

\obs{5}{
Over half of posters (77.9\%, 83.5\%, 70.6\% for Hash, Signature, and ZKP, respectively)
face implementation-related obstacles. These obstacles arise from identifying detailed implementation steps, reusing third-party templates, and operating crypto APIs.
}

While most questions focus on how to implement cryptographic tasks, we also found several posters (10.5\%, 5.8\%, and 8.8\%, for Hash, Signature, and ZKP, respectively) asking about the security of cryptographic practices.
They face difficulties in assessing the security of their implementations/designs and ask for assistance such as code review. %
In particular, 11 posters raise concerns about specific security vulnerabilities, \ie, \textit{weak random number generation} (7) and \textit{signature replay} (4). After examining these questions, we found that some posters were familiar with the concepts of these vulnerabilities, but encountered difficulties in identifying protective solutions or determining if their implementations were vulnerable.
Notably, compared to implementation issues, the number of security-related questions was relatively low.
It implies that while StackExchange posts provide valuable guidance for implementing cryptographic tasks, they may provide relatively less security-related guidance, such as security warnings and protective solutions, to developers.

\obs{6}{A small portion of posters (10.5\%, 5.8\%, and 8.8\%, for Hash, Signature, and ZKP, respectively) focus on making their implementations secure. The vulnerabilities they concern include weak random number generation and signature replay.}

\section{Answer to RQ3: The Practitioners' Perspective}
\label{sec:RQ2-Online-Survey}

The findings in Section~\ref{sec:RQ2-StackExchange-Posts} give us a preliminary understanding of developers' obstacles in on-chain cryptographic tasks. However, they may face two potential limitations. The first limitation stems from the fact that StackExchange questions may not fully represent the real-world obstacles encountered by developers. For example, the relatively few security-related questions on StackExchange do not definitively suggest that real-world practitioners seldom face security challenges. The second limitation pertains to the practitioners' perspective of the obstacles. While we have categorized potential obstacles, the underlying reasons behind these obstacles and possible solutions to mitigate them remain unexplored. 

To address these limitations, we conducted an online survey targeting real-world smart contract practitioners, to validate the identified obstacles, investigate the root causes, delve into practitioners' perceptions, and determine the type of support they require.

\subsection{Survey Design}
\

We followed Kitchenham and Pfleeger's instructions~\cite{kitchenham2008personal} for personal opinion surveys and designed an anonymous survey to increase response rates~\cite{tyagi1989effects}.
The following section provides a brief introduction to our survey questions. For the complete questionnaire, please refer to the online supplementary material~\cite{supplement}.

\myparagraph{Demographics}($Q_{1-4}$) We collected the following demographic information to understand respondents' background and distribution, and filter those who might not fully understand our survey.
\begin{itemize}
	\item Smart contract practitioner? Yes / No. ($Q_1$)
	\item Main role as a smart contract practitioner. Development / Testing / Project Management / Research / Other. ($Q_2$)
	\item Experience in years. Free-text. ($Q_3$)
	\item Current country of residence. Free-text. ($Q_4$)
\end{itemize}

\myparagraph{Cryptographic Practices}($Q_{5-6}$) This part aims to understand the participants' experience in on-chain cryptographic practices. 
We inquired about the crypto APIs ($Q_5$) they have used and whether they are familiar with their basic concepts ($Q_6$).

\myparagraph{Obstacles in Practitioners' Perspective}($Q_{7-10}$)
This part explores the difficulties of accomplishing cryptographic tasks from the practitioners' perspective.
We assessed the perceived difficulty in accomplishing cryptographic tasks compared to other tasks in smart contracts ($Q_{7}$) and asked the rationale behind their perceptions ($Q_{8}$).
Furthermore, we inquired about the obstacles contributing to the perceived difficulty, including the five identified obstacles and any others they faced ($Q_9$). We also asked participants how they acquired the knowledge for cryptographic tasks ($Q_{10}$).

\myparagraph{Template Usage}($Q_{11-12}$)
This part aims to understand the practitioners' perceptions of the usage of third-party cryptographic templates. We asked whether participants use templates to implement cryptographic tasks ($Q_{11}$), and whether existing templates sufficiently support their tasks ($Q_{12}$).

\myparagraph{Crypto API Design}($Q_{13-14}$)
This part investigates the practitioners' perceptions of Ethereum crypto APIs.
Specifically, we explored practitioners' perceptions of the functionality and usability of the current API design ($Q_{13}$). We also inquired about any cryptographic tasks they intend to accomplish but lack support from current APIs ($Q_{14}$).

\myparagraph{Security}($Q_{15-17}$) 
This part focuses on practitioners' perceptions of the security aspects of on-chain cryptographic practices.
We investigated the perceived difficulty associated with securing cryptographic tasks, compared to securing other tasks in smart contracts ($Q_{15}$) and asked for the rationale behind their perceptions ($Q_{16}$). Moreover, we inquired whether participants are familiar with crypto-related vulnerabilities listed in the Smart Contract Weakness Classification (SWC) List~\cite{swclist}, thereby exploring their understanding of crypto-specific security in smart contracts ($Q_{17}$).

\myparagraph{Resources and Tools}($Q_{18-19}$) 
To identify the types of support desired by practitioners,
we explored their perceptions of existing resources/tools ($Q_{18}$) and asked about any other types of support they may require ($Q_{19}$).

To reduce arbitrary responses caused by insufficient understanding of the questions, we provided an option of ``I do not understand this question'' for each multiple-choice question. For other questions, we explicitly stated that participants could skip them if they did not understand them.
Our survey was made available in both English and Chinese, since English is the most widely used language and Chinese has the largest number of speakers worldwide.
We carefully reviewed the two versions and guaranteed their consistency.

\subsubsection{Survey Validation}
We conducted a pilot survey with a small number of practitioners to obtain feedback on whether the questions are clear and easy to understand. \added{The participants included our collaborators and partners working in well-known blockchain companies.} Based on the feedback, we refined some questions for enhanced clarity without adding or removing any questions.
We also polished our translation to further reduce ambiguity between the two language versions of the survey.

\subsubsection{Participant Recruitment}
We adopted a non-probabilistic~\cite{gabor2007types} strategy for participant recruitment. Specifically, we conducted a keyword-based search for crypto-related smart contract repositories on Github, extracted their contributors' emails via the Github REST API~\cite{GithubAPI}, and sent the survey to them.
Our selected keywords encompass typical on-chain application scenarios of cryptography, including ``wallet'', ``token'', ``bridge'', and ``oracle'', as well as common primitives including ``hash'', ``signature'', and ``zkp''.
We also sent our survey to our industry partners working in well-known blockchain-related companies such as Meta and Alibaba.

We sent our survey to a total of 778 smart contract practitioners and received 78 responses from 21 countries. The response rate is 10.02\%, which is decent compared to previous studies~\cite{nadi2016jumping,bosu2019understanding}. 
We further excluded seven responses in which the respondents did not claim themselves as smart contract practitioners.
The main roles of the remaining 71 respondents are research (34, 47.9\%), development (32, 43.7\%), testing (2, 2.8\%), project management (1, 1.4\%), security audit (1, 1.4\%), and technical writing (1, 1.4\%). Their average years of experience
are 2.42 (min: 0.5, max: 7.0, median: 2.0, sd: 1.6).
\subsection{Results}

\label{sec:survey-result}
\

\myparagraph{Cryptographic Practices}
The top three most common crypto APIs used by participants are \KECCAK, \SHATWO, and \Ecrecover, utilized by 84.5\%, 67.6\%, and 57.7\% of them, respectively.
Most of the participants demonstrated familiarity with the concepts of Hash and Signature primitives (98.6\% and 93.0\%, respectively), while only 50.7\% of participants were familiar with ZKP.
Interestingly, although 50.7\% of the participants claimed familiarity with ZKP, only about 10\% of participants have used ZKP-related APIs (\ECADD (12.7\%), \ECMUL (11.3\%), and \ECPAIRING (12.7\%)) in practice, suggesting a gap between conceptual knowledge and practical implementation.

\myparagraph{Obstacles in Practitioners' Perspective}
We explored the perceived challenges in accomplishing on-chain cryptographic tasks.
We found that 54.8\% of participants believed that accomplishing cryptographic tasks is more challenging than accomplishing other tasks such as business logic. 
An additional 14.5\% of participants believed that the challenges are comparable in magnitude, but cryptographic practices involve additional challenges arising from different aspects.
Such results indicate that the technical stack required for cryptographic tasks may differ from that of common business logic, making it challenging for regular smart contract developers to handle cryptographic tasks.

\obs{7}{In comparison to other programming tasks in smart contracts, 69.3\% of participants believe that accomplishing cryptographic tasks presents additional challenges to them.} 

To understand the root causes of these perceived challenges, we asked participants about the specific obstacles they encountered.
Specifically, we included the five categories of obstacles identified from StackExchange posts in RQ2 as choices, while also allowing adding new obstacles through the ``Other'' choice.
As shown in Table~\ref{tab:survey-obstacle}, real-world practitioners indeed face these five categories of obstacles.
Additionally, none of the participants mentioned new obstacles, indicating the completeness of our classification.

The top two common obstacles reported by 64 respondents are \textit{Roadmap Identification} and \textit{security}.
In particular, 57.8\% of participants face obstacles in identifying the roadmap to implement their tasks, even before they actually begin the programming process.
Such developers require more high-level guidance, such as task-based templates and easy-to-understand tutorials.
Additionally, even after implementing their tasks, 56.3\% of participants confront obstacles in evaluating the security of their implementation.

\obs{8}{
The most frequently faced obstacles is identifying detailed steps to implement specific tasks (faced by 57.8\% of participants), followed by evaluating the security of the implementations (faced by 56.3\%).
}

To delve deeper into practitioners' perception, we conducted a free-text question, asking participants what makes cryptographic tasks challenging.
While the majority (84.6\%, 33/39) of responses could be categorized into the five obstacles listed in Table~\ref{tab:survey-obstacle}, we found six interesting comments that attribute obstacles to the design of Ethereum crypto APIs.
For example, respondents pointed out that ``\textit{Solidity lacks native cryptographic libraries}'', ``\textit{Cryptographic functions provided in Solidity are quite complex to use}'', 
``\textit{(cryptographic tasks) often require inline assembly to interact with precompiles}'', suggesting that the current crypto APIs in Solidity are too low-level for them to use.
Additionally, two of the six participants also highlighted that the high gas cost in Solidity hinders their cryptographic practices.

\obs{9}{Several participants attribute obstacles they encountered to the design of Ethereum crypto APIs.
They perceive current APIs as too low-level to directly interact with.}

To explore the relationship among different obstacles, we conducted a Chi-squared test~\cite{holt1980chi} to examine the potential correlation among the five types of obstacles. We observed statistically significant correlations (p-value < 0.05) between two pairs of obstacles, \ie, \textit{API usage} and \textit{Template usage} (p-value = 0.042), \textit{Knowledge} and \textit{Roadmap Identification} (p-value = 0.021), indicating their relevance in the practitioners' perspective.
The former correlation suggests that: %
since templates are fixed combinations of APIs, the obstacles in using templates might stem from the obstacles in using the APIs.
For example, OpenZepplin's ECDSA template~\cite{openzepplinECDSA} internally calls the \KECCAK and \Ecrecover API to implement ECDSA signature verification. Therefore, using the template might require the understanding of the \Ecrecover API.
The latter correlation suggests that having insufficient knowledge about cryptography or blockchain can directly prevent developers from identifying the specific implementation steps for their tasks.
For example, if one does not understand the hash-then-sign paradigm~\cite{mironov2006collision} of signatures, they may not know the correct sequence of API calls for signature verification either.
All other pairs are weakly correlated, indicating that they are distinct obstacles from the practitioners' perspective.

\begin{table}[]
	\centering
	\caption{Obstacles faced by participants}
  \vspace{-1em}
	\label{tab:survey-obstacle}
	\begin{tabular}{lccc}
		\toprule
		\textbf{Obstacles}& \textbf{\# Participants} & \textbf{Ratio} & \textbf{Rank}  \\ \midrule
		Roadmap Identification                   & 37    & 57.8\%        & 1                 \\
		\rowcolor{gray!30} 	Security & 36    & 56.3\%        & 2        \\
		Knowledge              & 26   & 40.6\%        & 3                 \\
		\rowcolor{gray!30} Template Usage                   & 20   & 31.3\%       & 4                    \\
		API Usage                     & 14    & 21.9\%            & 5                \\  \bottomrule
	\end{tabular}

\end{table}

\myparagraph{Crypto API Design}
To explore the the participants' perceptions of current Ethereum crypto APIs, we asked them to rate the APIs on a 5-point Likert scale~\cite{joshi2015likert} (\emph{very bad, bad, neutrality, good, very good}).
Specifically, we provided two aspects of the Ethereum crypto APIs for rating: (1) Functionality: whether the functionalities of current Ethereum crypto APIs can support all cryptographic tasks developers need? 
(2) Usability: whether these APIs are easy to learn and use, even for non-expert developers?
55.9\% and 47.1\% of participants rated the functionality and usability of current crypto APIs as \emph{good} or \emph{very good}, respectively.
Furthermore, only 36.8\% of participants rated both aspects as \emph{good} or \emph{very good}, demonstrating a large room for improvement of the current crypto APIs.

To investigate whether the functionality of Ethereum crypto APIs fully meets real-world demands,
we inquired if there were any tasks practitioners wished to accomplish but were not supported by current APIs.
Ten respondents indicated that current APIs lack the functionality they need.
Six of them explicitly requested support for other elliptic curves in addition to the current alt\_bn128 curve supported by \ECADD, \ECMUL, and \ECPAIRING.
The needed curves included BLS12-381~\cite{barreto2003constructing}, Pasta~\cite{pasta}, and other pairing-friendly curves~\cite{barreto2005pairing}, which provide necessary operations for several currently unavailable tasks, \eg, recursive zero-knowledge proof systems.
Four respondents requested underlying support for cryptographic schemes, including BLS signature~\cite{boneh2001short}, Schnorr signature~\cite{schnorr1990efficient}, and BBS group signature~\cite{boneh2004short}, which also require adding additional elliptic curve operations.
For example, public key aggregation of BLS signature is unsupported by current APIs~\cite{blssignature}.
Although several Ethereum Improvement Proposals (EIPs)~\cite{eip-1895,eip-2537} proposed adding new elliptic curves to Ethereum, none of them has been adopted.

\obs{10}{ 
Only 36.8\% of participants expressed satisfaction with current Ethereum crypto APIs in terms of functionality and usability.
Current APIs lack support for several cryptographic tasks, especially those requiring other elliptic curves.}

\myparagraph{Template Usage}
To explore how third-party cryptographic templates fit into real-world practices, we asked participants whe-ther they use existing templates or implement cryptographic tasks directly based on underlying APIs.
The results showed that only 4.6\% of participants always choose to implement tasks directly based on Ethereum crypto APIs, while 95.4\% tend to use templates for some or all of the cryptographic tasks they implement.
Furthermore, we inquired if there were any cryptographic tasks that participants desired to accomplish but lacked available templates.
\added{The responses included Zero-knowledge proof schemes (5 responses), Verkle proof~\cite{kuszmaul2019verkle} (1 response), and cryptographic accumulator~\cite{ozcelik2021overview} (1 response).}
Compared to tasks such as ECDSA signature verification, these tasks are relatively less used and have fewer code examples for reference. Therefore, they might be the tasks that require templates the most.

\obs{11}{ 
Third-party templates can assist developers in implementing cryptographic tasks. 
Participants desire more templates, especially for emerging tasks such as zero-knowledge proof.
}

\begin{table}[]
	\centering
	\caption{Participants' perceptions of current resources / tools}
 \setlength\tabcolsep{3pt}%
 \vspace{-1em}
	\label{tab:resource}
	\begin{tabular}{lccc}
		\toprule
		\textbf{Resources / Tools}& \textbf{Distribution} & \textbf{Average}  & \textbf{ Score$\ge$4} \\ \midrule
		Official Document                   & \raisebox{-.\height}{\includegraphics[width=.3\linewidth]{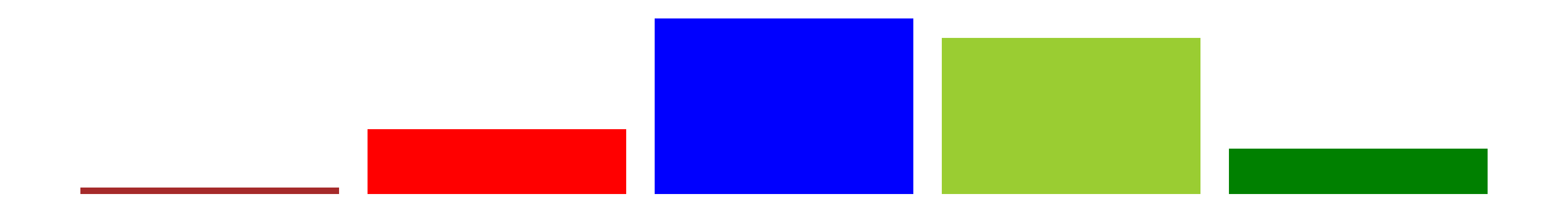}}    &  3.38 & 44.9\% \\ 
		Audited Templates & \raisebox{-.\height}{\includegraphics[width=.3\linewidth]{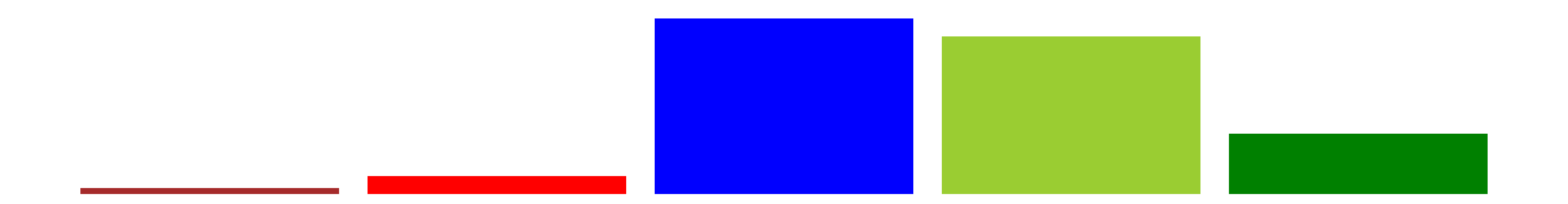}}     & 3.59   & 52.2\%   \\
		Testing Tools    & \raisebox{-.\height}{\includegraphics[width=.3\linewidth]{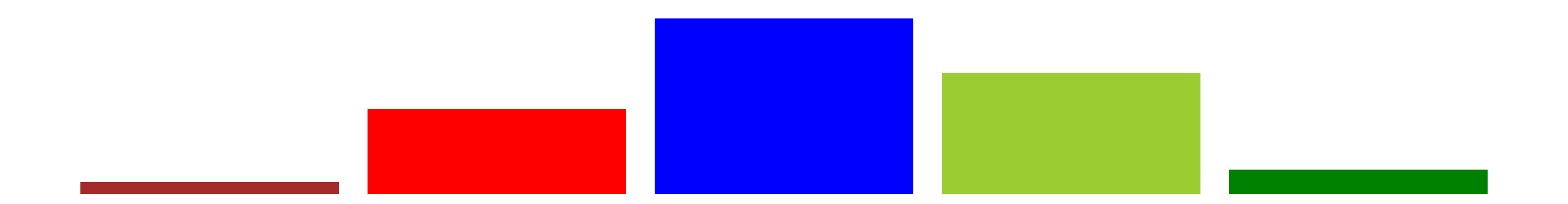}}      & 3.14  & 34.8\%    \\ 
		Audit Tools    & \raisebox{-.\height}{\includegraphics[width=.3\linewidth]{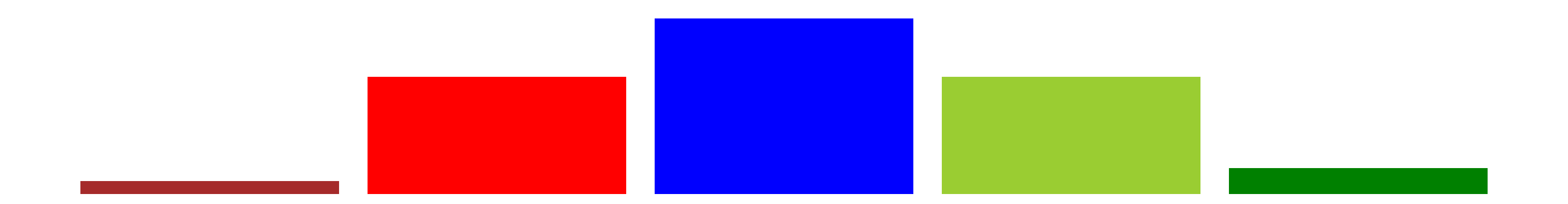}}   & 3.06   & 31.9\%     \\ 
		 \bottomrule
	\end{tabular}
 \vspace{-1em}

\end{table}

\myparagraph{Security}
This part investigated the practitioners' perceptions of the security of cryptographic practices and explored their understanding of crypto-related vulnerabilities.
We found that compared to other common tasks in smart contracts, 64.1\% of participants perceive additional challenges in securing their cryptographic tasks.
To attribute these challenges, we further questioned the underlying reasons.
Out of the 32 participants who provided feedback, 
19 (59.4\%) participants mentioned that cryptographic practices require cryptographic expertise and low-level operations, which bring more security issues inherently. For example, the respondents mentioned
``\textit{Easy to get in a pitfall with bytes/bits around you}'', ``\textit{Only understanding cryptography can secure implementations while cryptography is not easy to learn}''. Additionally, eight (25.0\%) participants suggested that it is difficult to detect vulnerabilities in cryptographic implementations. For example, respondents mentioned that ``\textit{(cryptographic tasks are) impossible to verify, almost always probabilistic, no way to test}'' and ``\textit{No smart contract security analyzer supports the detection of cryptographic-related bugs}''.
It demonstrates the necessity to
inform developers of the security implications behind the Ethereum crypto APIs and provide improved techniques to detect security vulnerabilities.

\obs{12}{64.1\% of participants facing additional challenges in securing their cryptographic tasks. They attribute these challenges to the complexity of cryptographic operations and the lack of detection methods for crypto-specific vulnerabilities.
}

Furthermore, we inquired about the practitioners' familiarity with crypto-related vulnerabilities.
In line with previous studies~\cite{piantadosi2023detecting,song2022esbmc,sendner2023smarter,ivanov2023txt}, we focused on vulnerabilities in the Smart Contract Weakness Classification (SWC) list~\cite{swclist}.
Analogous to the Common Weakness Enumeration (CWE) list for vulnerabilities in traditional software, the SWC list documents security vulnerabilities in Ethereum smart contracts.
Specifically, it included the following five types of crypto-related vulnerabilities:
\begin{itemize}[leftmargin=*]
\item SWC-117: Signature Malleability
\item SWC-120: Weak Sources of Randomness from Chain Attributes
\item SWC-121: Missing Protection against Signature Replay Attacks
\item SWC-122: Lack of Proper Signature Verification
\item SWC-133: Hash Collisions With Multiple Variable Length Arguments
\end{itemize}

The results showed that 30.9\% of participants know all five types of vulnerabilities, while 11.6\% of participants know none of them. On average, participants know 2.9 types of vulnerabilities (median: 3.0, sd: 1.8), indicating a moderate and improvable level of knowledge regarding crypto-related vulnerabilities.
Among the five types of vulnerabilities, \textit{Weak Sources of Randomness from Chain Attributes} and \textit{Missing Protection against Signature Replay Attacks} are most familiar to participants, both of which are known by 62.3\% of participants.
Such understanding is essential for employing standard protection and preventing insecure cryptographic practices.

\myparagraph{Resources and Tools}
We examined whether the community support meets the needs of practitioners.
Specifically, we focused on four types of resources and tools that are commonly used by contract developers~\cite{chen2021maintenance,wan2021smart}, \ie, documentation, templates, testing tools, and audit tools, and asked participants to rate them on a 5-point Likert scale (\emph{very bad, bad, neutrality, good, very good}).
As shown in Table~\ref{tab:resource}, the average scores for all resources/tools were below 4 (\emph{good}).
In particular, the testing and auditing tools received the lowest scores, suggesting that current tools for general programming practices might need to be fine-tuned to better support cryptographic tasks.
In addition, participants mentioned the need for other tools,
such as language server protocol (LSP), editor plugin, and domain-specific language (DSL) for programming cryptographic tasks.

\obs{13}{
Only 34.8\% and 31.9\% of participants are satisfied with existing testing tools and audit tools respectively, demonstrating a lack of tools to test and secure cryptographic practices.
}

\section{Discussion}

In this section, we discuss the implications of our findings and point out future directions to improve the Ethereum crypto APIs, facilitate the security of cryptographic practices, and improve third-party templates and tools.

\subsection{Improving Ethereum Crypto APIs}
\

Our survey results indicate the need to improve the functionality and usability of current Ethereum APIs.

\myparagraph{Usability}
Previous studies on traditional cryptographic libraries~\cite{nadi2016jumping, patnaik2019usability} highlighted that it is essential to hide unnecessary underlying details and provide high-level interfaces for cryptographic tasks.
For example, Java Cryptography Architecture~\cite{weiss2004java} provides ready-to-use APIs for high-level cryptographic tasks. Developers can use statements like \emph{Signature.getInstance(``SHA256-withECDSA'')} to implement signatures in Java, without operating low-level cryptographic operations themselves.
To alleviate developers from managing low-level operations, future work could focus on \textbf{supporting built-in cryptographic libraries in Solidity}.
In fact, four crypto APIs (\KECCAK, \Ecrecover, \SHATWO, and \RIPEMD) have already become built-in functions in Solidity.
Developers can use keywords like ``\emph{ecrecover}'' to invoke these functions, and the Solidity compiler will automatically transform them into calls to the corresponding crypto APIs.
However, the challenging task of composing these low-level APIs to accomplish higher-level tasks still falls on developers, requiring further exploration and improvement.

\myparagraph{Functionality}
Our studies demonstrate that the limited functionalities of current APIs hindered the applications of several promising tasks, \eg, recursive zero-knowledge proofs (\textit{Observation 10}).
Notably, in traditional software, developers have the flexibility to explore alternative libraries or even implement missing functionality themselves.
However, in smart contracts, the high gas cost of cryptographic operations makes these system-level crypto APIs the most practical option for developers.
Therefore, to avoid fundamentally impeding the application of emerging cryptographic tasks, it is necessary to introduce new crypto APIs required by real-world practices.

\added{Crypto operations mentioned by our real-world survey participants, such as paring-friendly elliptic curves could be promising candidates for improving the functionality of Ethereum cryptographic APIs. In addition to our survey, there is other real-world evidence supporting the inclusion of these candidates. For example, Go-Ethereum~\cite{goethereum} have incorporated BLS12-381~\cite{barreto2003constructing} into EVM for testing purpose, and several Solidity libraries~\cite{curveSol,pastaSol} have implemented elliptic curves like Pasta~\cite{pasta}, even with impractical gas cost. However, given the high cost to add precompiled contracts and the backward compatibility requirements, the addition of these candidates might still need to be further evaluated within more detailed contexts.}
Future work could put more efforts into \textbf{collecting real-world evidences of adding these candidates}, \eg, conducting empirical research on the real-world requirements of Ethereum cryptographic APIs.

\subsection{The Security of Cryptographic Practices}
The security practices for on-chain cryptographic tasks are still in the early stages of development, with deficiencies in both knowledge and tools.
For example, while digital signatures have become a common cryptographic task in smart contracts, 37.7\% of developers are unfamiliar with potential vulnerabilities related to signature replay attacks (SWC-121).
The insecure practices caused by such insufficient understanding have caused real-world security issues~\cite{defcon}.
Therefore, it is necessary to \textbf{inform developers about the security implications behind Ethereum crypto APIs} and \textbf{provide guidelines to prevent common misuses}.

Furthermore, compared to extensive research~\cite{zhang2022automatic} focusing on traditional cryptographic libraries, 
the security of Ethereum crypto APIs has not attracted enough attention.
While crypto API misuses have been established as common causes of security vulnerabilities~\cite{zhang2022automatic,gu2019empirical,egele2013empirical}, what crypto API misuse exists in real-world Ethereum cryptographic practices still lacks systematic analysis.
In addition, existing security analysis tools for smart contracts also lack support for crypto-specific vulnerabilities.
For example, a recent work~\cite{zhang2023demystifying} conducted a literature review of papers published on top-tier Software Engineering, Security, and Programming Language venues from 2017 to 2022 and documented 17 categories of vulnerabilities that can be detected by existing security tools. However, only the "\textit{weak sources of randomness}" among them is related to cryptographic practices,
indicating a lack of tools to detect other crypto-specific vulnerabilities, such as those listed in the SWC list~\cite{swclist}.
Future studies could focus on \textbf{characterizing Ethereum crypto API misuses in real-world practices} and \textbf{incorporating crypto-specific vulnerabilities into existing security analysis tools}.

\subsection{Templates and Tools for Implementing Cryptographic Tasks}
The complexity of cryptographic implementations often requires developers to rely on templates and tools to fulfill their development needs. For example,
95.4\% of the survey participants use templates to accomplish some or all of their cryptographic tasks.
However, the feedback from the participants suggests a need for improvement in terms of the usability and functionality of existing templates.
For example, OpenZeppelin~\cite{openzepplin}, one of the most popular template libraries, only provides templates for ECDSA signatures and Merkle proofs, while lacking support for other tasks, such as the zero-knowledge proof mentioned by participants.
Furthermore, 32.8\% of the survey participants encountered difficulties in using templates.
Future studies could focus on \textbf{improving the usability of existing templates} and \textbf{introducing new templates to cover emerging tasks}.
Furthermore, several previous studies~\cite{steffen2019zkay, steffen2022zeestar} and industrial solutions~\cite{belles2022circom, eberhardt2018zokrates} have explored using domain-specific languages and code generation tools to automate the code generation process for zero-knowledge proofs.
It could be valuable to further \textbf{explore the application of code generation tools} for other on-chain cryptographic tasks.

\section{Threats to Validity}
\subsection{Internal Validity}

\

Section~\ref{sec:RQ1} and Section~\ref{sec:RQ2-StackExchange-Posts} relied on manual analysis of smart contracts and StackExchange posts, respectively.
Therefore, these analysis results might be inherently subjective.
To mitigate the subjective biases and strengthen the reliability of our results, we implemented a dual-author approach to every analysis, ensuring that any disagreements were thoroughly resolved.
Notably, all authors involved in these tasks have more than three years of experience in smart contract research.
In addition, due to the large datasets and time-consuming analysis process, we utilized random sampling to make manual analysis feasible.
To mitigate potential sampling bias, the sampling size and ratio for each study are meticulously determined, based on an extensive analysis of previous studies~\cite{nadi2016jumping,chen2020defining,yang2023definition} that examined analogous data sources.
We also published the labeled dataset and results in our online supplement materials to facilitate replication or further analysis.

Since our study focused on a relatively specific topic,
(cryptographic practices on Ethereum),
the survey participants in Section~\ref{sec:RQ2-Online-Survey} might have insufficient understanding of the questions, which could lead to irrational responses. To reduce the impact, we included an ``I don’t understand'' option and excluded such responses from analysis.
We also explicitly requested developers to skip questions they found incomprehensible.
\added{Several questions in our survey use subjective words (\eg, the term “more challenging” in $Q_7$) to explore participants’ perceptions, which could potentially influence the responses. To mitigate the impact of such suggestiveness on our results, we have implemented several measures, including asking for the rationales behind participants’ choices and offering additional choices beyond a simple “\textit{Yes}” or “\textit{No}” to avoid  forced decision-making. For example, in $Q_7$, participants who chose “\textit{Yes}” or “\textit{They are basically the same, but lie in different aspects}” were requested to explicitly explain their choices in $Q_8$. 39 out of 45 (86.7\%) of participants provided detailed rationales, which potentially validates their initial responses to $Q_7$.}
Additionally, we adopted a non-probabilistic participant recruitment process for the survey. Although it may result in a pool that is not
fully representative of all practitioners, we have tried our best to cover the whole population.
Our survey included participants from 21 countries, all with varied roles and experiences.

\subsection{External Validity}
Solidity is an ever-evolving language,
with new crypto APIs introduced through periodic hard-forks~\cite{hardfork}.
For example, the Cancun upgrade~\cite{cancun} will add a new precompiled contract~\cite{eip-4844} to verify KZG proofs~\cite{kate2010constant} in smart contracts.
Although our current results cannot cover crypto APIs introduced in the future, our method, findings, and datasets can be extended with minor efforts to incorporate them.

\section{Related Work}
\subsection{General Programming Practices in Smart Contracts}

The rapid development of smart contracts has motivated a substantial body of empirical studies on programming practices in smart contracts.
They covered the implementation~\cite{bosu2019understanding,zou2019smart,liao2022large,liu2021characterizing}, security~\cite{chen2020defining,wan2021smart}, and many other aspects~\cite{bartoletti2017empirical,ajienka2020empirical} of programming practices.
For example, Bosu~\etal~\cite{bosu2019understanding} explored the differences between the implementation of smart contracts and traditional software. They found that 93\% of developers believe that smart contract development differs from traditional software development, primarily due to its higher emphasis on security and reliability.
Chen~\etal~\cite{chen2021maintenance} studied smart contract maintenance-related concerns based on the survey feedback and literature review.
Wan~\etal~\cite{wan2021smart} conducted interviews and surveys to investigate how real-world practitioners build security into smart contract development.
They found that 85\% and 69\% of the practitioners recognize the importance of security and privacy in smart contracts respectively. They commonly re-use templates and employ security tools to enhance the security.
However, to the best of our knowledge, these studies mainly focused on general programming tasks, while lack analysis of crypto-specific practices.

\subsection{Cryptographic Practices in Traditional Software}

Several studies~\cite{egele2013empirical,nadi2016jumping,hazhirpasand2020java} have been conducted to characterize cryptographic practices in traditional software.
Hazhirpasand~\etal~\cite{hazhirpasand2020java} collected 489 Github Java projects using Java Cryptography Architecture (JCA) and analyzed the uses and misuses of APIs in them.
They found that 99.8\% projects using JCA APIs contain at least one misuse.
Nadi~\etal\cite{nadi2016jumping} found that JCA is the most used cryptographic library, and securing connections, authenticating user logins, and encrypting files are the top three tasks developers need.
Since Ethereum crypto APIs are different from traditional crypto libraries in both design and usage, the results of these studies might not be applicable to Ethereum.
\added{Notably, while the majority of our results (\textit{Observation 1-3, 10-13}) are specific to the Ethereum context, some observations such as the categorization of obstacles in \textit{Observation 4-9}, might also be generally applicable. They can validate previous results and provide new evidence of their relevance within the Ethereum context.}

\subsection{The Usability of Crypto APIs}

Several previous studies have focused on the usability of crypto APIs to mitigate the obstacles developers face.
Green and Smith~\cite{green2016developers} proposed ten principles for creating usable and secure crypto APIs, 
\eg, \textit{integrate crypto functionality into APIs so regular developers don’t have to interact with crypto APIs in the first place.}.
These principles have been used in several following studies~\cite{acar2017comparing,mindermann2018usable,patnaik2019usability}.
Patnaik~\etal~\cite{patnaik2019usability} identified 16 usability issues of seven existing crypto libraries by analyzing StackOverflow questions and provided evidence to validate Green and Smith's heuristics~\cite{green2016developers}.
Nadi~\etal~\cite{nadi2016jumping} analyze why java developer struggle with crypto APIs by analyzing StackOverflow questions and surveying developers.
Acar~\etal~\cite{acar2017comparing} conducted a controlled experiment with Python developers to evaluate the usability of five Python cryptographic libraries. They found that libraries designed for simplicity can reduce the decision space of developers and offer security benefits.
They also suggested ensuring support for common tasks and providing accessible documentation to developers.
Kai~\etal~\cite{mindermann2018usable} identified seven major cryptographic libraries in Rust based on a search on Github and other sources, and examined their usability through controlled experiments.
Different from these previous studies, we focus on the usability of crypto APIs in the specific context of Ethereum smart contracts.
Notably, we not only analyzed usability issues from a single information source, such as StackExchange posts, but also conducted a survey with smart contract practitioners to validate our findings and explore their perceptions.

\section{Conclusion}

We conducted the first empirical study on cryptographic practices in Ethereum smart contracts, through examining 91,484,856 Ethereum transactions and 500 smart contracts, 483 StackExchange posts, and survey input from 78 smart contract practitioners.
Our results showed that while Ethereum crypto APIs enable prevalent and diverse on-chain cryptographic tasks, they also pose obstacles for developers.
We identified five categories of obstacles from StackExchange posts and conducted an online survey to gain insights from practitioners. Our findings revealed that 57.8\% of practitioners encounter obstacles in identifying the detailed steps to implement specific tasks, and 56.3\% face difficulties in evaluating the security of their implementations. 
The feedback from the participants highlighted the gap between low-level crypto APIs and high-level cryptographic tasks
and demonstrated the need for improved API functionality and usability, task-based templates, and effective assistance tools.
Based on these findings, we provided practical implications to the API designers and template/tool providers and outlined possible directions for future research.

\bibliographystyle{ACM-Reference-Format}
\bibliography{main}


\begin{thebibliography}{107}


\ifx \showCODEN    \undefined \def \showCODEN     #1{\unskip}     \fi
\ifx \showDOI      \undefined \def \showDOI       #1{#1}\fi
\ifx \showISBNx    \undefined \def \showISBNx     #1{\unskip}     \fi
\ifx \showISBNxiii \undefined \def \showISBNxiii  #1{\unskip}     \fi
\ifx \showISSN     \undefined \def \showISSN      #1{\unskip}     \fi
\ifx \showLCCN     \undefined \def \showLCCN      #1{\unskip}     \fi
\ifx \shownote     \undefined \def \shownote      #1{#1}          \fi
\ifx \showarticletitle \undefined \def \showarticletitle #1{#1}   \fi
\ifx \showURL      \undefined \def \showURL       {\relax}        \fi
\providecommand\bibfield[2]{#2}
\providecommand\bibinfo[2]{#2}
\providecommand\natexlab[1]{#1}
\providecommand\showeprint[2][]{arXiv:#2}

\bibitem[Acar et~al\mbox{.}(2017)]%
        {acar2017comparing}
\bibfield{author}{\bibinfo{person}{Yasemin Acar}, \bibinfo{person}{Michael
  Backes}, \bibinfo{person}{Sascha Fahl}, \bibinfo{person}{Simson Garfinkel},
  \bibinfo{person}{Doowon Kim}, \bibinfo{person}{Michelle~L Mazurek}, {and}
  \bibinfo{person}{Christian Stransky}.} \bibinfo{year}{2017}\natexlab{}.
\newblock \showarticletitle{Comparing the usability of cryptographic apis}. In
  \bibinfo{booktitle}{\emph{2017 IEEE Symposium on Security and Privacy (SP)}}.
  IEEE, \bibinfo{pages}{154--171}.
\newblock


\bibitem[Ajienka et~al\mbox{.}(2020)]%
        {ajienka2020empirical}
\bibfield{author}{\bibinfo{person}{Nemitari Ajienka}, \bibinfo{person}{Peter
  Vangorp}, {and} \bibinfo{person}{Andrea Capiluppi}.}
  \bibinfo{year}{2020}\natexlab{}.
\newblock \showarticletitle{An empirical analysis of source code metrics and
  smart contract resource consumption}.
\newblock \bibinfo{journal}{\emph{Journal of Software: Evolution and Process}}
  \bibinfo{volume}{32}, \bibinfo{number}{10} (\bibinfo{year}{2020}),
  \bibinfo{pages}{e2267}.
\newblock


\bibitem[Alex et~al\mbox{.}(2020)]%
        {eip-2537}
\bibfield{author}{\bibinfo{person}{Vlasov Alex}, \bibinfo{person}{Olson Kelly},
  {and} \bibinfo{person}{Stokes Alex}.} \bibinfo{year}{2020}\natexlab{}.
\newblock \bibinfo{title}{EIP-1895: Support for an Elliptic Curve Cycle}.
\newblock
\newblock
\urldef\tempurl%
\url{https://eips.ethereum.org/EIPS/eip-1895}
\showURL{%
Retrieved June 17, 2023 from \tempurl}


\bibitem[Anonymous(2023)]%
        {supplement}
\bibfield{author}{\bibinfo{person}{Anonymous}.}
  \bibinfo{year}{2023}\natexlab{}.
\newblock \bibinfo{title}{Online Supplement Material}.
\newblock
\newblock
\urldef\tempurl%
\url{https://zenodo.org/records/10074040}
\showURL{%
\tempurl}


\bibitem[Bai(2018)]%
        {defcon}
\bibfield{author}{\bibinfo{person}{Zhenxuan Bai}.}
  \bibinfo{year}{2018}\natexlab{}.
\newblock \bibinfo{title}{You may pay more than you can imagine}.
\newblock
\newblock
\urldef\tempurl%
\url{https://github.com/nkbai/defcon26/tree/master/docs}
\showURL{%
\tempurl}


\bibitem[Barreto et~al\mbox{.}(2003)]%
        {barreto2003constructing}
\bibfield{author}{\bibinfo{person}{Paulo~SLM Barreto}, \bibinfo{person}{Ben
  Lynn}, {and} \bibinfo{person}{Michael Scott}.}
  \bibinfo{year}{2003}\natexlab{}.
\newblock \showarticletitle{Constructing elliptic curves with prescribed
  embedding degrees}. In \bibinfo{booktitle}{\emph{Security in Communication
  Networks: Third International Conference, SCN 2002 Amalfi, Italy, September
  11--13, 2002 Revised Papers 3}}. Springer, \bibinfo{pages}{257--267}.
\newblock


\bibitem[Barreto and Naehrig(2005)]%
        {barreto2005pairing}
\bibfield{author}{\bibinfo{person}{Paulo~SLM Barreto} {and}
  \bibinfo{person}{Michael Naehrig}.} \bibinfo{year}{2005}\natexlab{}.
\newblock \showarticletitle{Pairing-friendly elliptic curves of prime order}.
  In \bibinfo{booktitle}{\emph{International workshop on selected areas in
  cryptography}}. Springer, \bibinfo{pages}{319--331}.
\newblock


\bibitem[Bartoletti and Pompianu(2017)]%
        {bartoletti2017empirical}
\bibfield{author}{\bibinfo{person}{Massimo Bartoletti} {and}
  \bibinfo{person}{Livio Pompianu}.} \bibinfo{year}{2017}\natexlab{}.
\newblock \showarticletitle{An empirical analysis of smart contracts:
  platforms, applications, and design patterns}. In
  \bibinfo{booktitle}{\emph{Financial Cryptography and Data Security: FC 2017
  International Workshops, WAHC, BITCOIN, VOTING, WTSC, and TA, Sliema, Malta,
  April 7, 2017, Revised Selected Papers 21}}. Springer,
  \bibinfo{pages}{494--509}.
\newblock


\bibitem[Beiko(2023)]%
        {cancun}
\bibfield{author}{\bibinfo{person}{Tim Beiko}.}
  \bibinfo{year}{2023}\natexlab{}.
\newblock \bibinfo{title}{Cancun Network Upgrade Meta Thread}.
\newblock
\newblock
\urldef\tempurl%
\url{https://ethereum-magicians.org/t/cancun-network-upgrade-meta-thread/12060/1}
\showURL{%
Retrieved June 17, 2023 from \tempurl}


\bibitem[Bell{\'e}s-Mu{\~n}oz et~al\mbox{.}(2022)]%
        {belles2022circom}
\bibfield{author}{\bibinfo{person}{Marta Bell{\'e}s-Mu{\~n}oz},
  \bibinfo{person}{Miguel Isabel}, \bibinfo{person}{Jose~Luis Mu{\~n}oz-Tapia},
  \bibinfo{person}{Albert Rubio}, {and} \bibinfo{person}{Jordi Baylina}.}
  \bibinfo{year}{2022}\natexlab{}.
\newblock \showarticletitle{Circom: A Circuit Description Language for Building
  Zero-knowledge Applications}.
\newblock \bibinfo{journal}{\emph{IEEE Transactions on Dependable and Secure
  Computing}} (\bibinfo{year}{2022}).
\newblock


\bibitem[Belling(2018)]%
        {eip-1895}
\bibfield{author}{\bibinfo{person}{Alexandre Belling}.}
  \bibinfo{year}{2018}\natexlab{}.
\newblock \bibinfo{title}{EIP-1895: Support for an Elliptic Curve Cycle}.
\newblock
\newblock
\urldef\tempurl%
\url{https://eips.ethereum.org/EIPS/eip-1895}
\showURL{%
Retrieved June 17, 2023 from \tempurl}


\bibitem[Bloemen et~al\mbox{.}(2017)]%
        {eip-712}
\bibfield{author}{\bibinfo{person}{Remco Bloemen}, \bibinfo{person}{Leonid
  Logvinov}, {and} \bibinfo{person}{Jacob Evans}.}
  \bibinfo{year}{2017}\natexlab{}.
\newblock \bibinfo{title}{EIP-712: Typed structured data hashing and signing}.
\newblock
\newblock
\urldef\tempurl%
\url{https://eips.ethereum.org/EIPS/eip-712}
\showURL{%
Retrieved June 17, 2023 from \tempurl}


\bibitem[Boneh et~al\mbox{.}(2004)]%
        {boneh2004short}
\bibfield{author}{\bibinfo{person}{Dan Boneh}, \bibinfo{person}{Xavier Boyen},
  {and} \bibinfo{person}{Hovav Shacham}.} \bibinfo{year}{2004}\natexlab{}.
\newblock \showarticletitle{Short Group Signatures}. In
  \bibinfo{booktitle}{\emph{Advances in Cryptology -- CRYPTO 2004}},
  \bibfield{editor}{\bibinfo{person}{Matt Franklin}} (Ed.).
  \bibinfo{publisher}{Springer Berlin Heidelberg}, \bibinfo{address}{Berlin,
  Heidelberg}, \bibinfo{pages}{41--55}.
\newblock


\bibitem[Boneh et~al\mbox{.}(2001)]%
        {boneh2001short}
\bibfield{author}{\bibinfo{person}{Dan Boneh}, \bibinfo{person}{Ben Lynn},
  {and} \bibinfo{person}{Hovav Shacham}.} \bibinfo{year}{2001}\natexlab{}.
\newblock \showarticletitle{Short signatures from the Weil pairing}. In
  \bibinfo{booktitle}{\emph{Advances in Cryptology—ASIACRYPT 2001: 7th
  International Conference on the Theory and Application of Cryptology and
  Information Security Gold Coast, Australia, December 9--13, 2001 Proceedings
  7}}. Springer, \bibinfo{pages}{514--532}.
\newblock


\bibitem[Bosu et~al\mbox{.}(2019)]%
        {bosu2019understanding}
\bibfield{author}{\bibinfo{person}{Amiangshu Bosu}, \bibinfo{person}{Anindya
  Iqbal}, \bibinfo{person}{Rifat Shahriyar}, {and} \bibinfo{person}{Partha
  Chakraborty}.} \bibinfo{year}{2019}\natexlab{}.
\newblock \showarticletitle{Understanding the motivations, challenges and needs
  of blockchain software developers: A survey}.
\newblock \bibinfo{journal}{\emph{Empirical Software Engineering}}
  \bibinfo{volume}{24}, \bibinfo{number}{4} (\bibinfo{year}{2019}),
  \bibinfo{pages}{2636--2673}.
\newblock


\bibitem[Buterin(2017)]%
        {eip-198}
\bibfield{author}{\bibinfo{person}{Vitalik Buterin}.}
  \bibinfo{year}{2017}\natexlab{}.
\newblock \bibinfo{title}{EIP-198: Big integer modular exponentiation}.
\newblock
\newblock
\urldef\tempurl%
\url{https://eips.ethereum.org/EIPS/eip-198}
\showURL{%
Retrieved June 17, 2023 from \tempurl}


\bibitem[Buterin(2018)]%
        {eip-1014}
\bibfield{author}{\bibinfo{person}{Vitalik Buterin}.}
  \bibinfo{year}{2018}\natexlab{}.
\newblock \bibinfo{title}{EIP-1014: Skinny CREATE2}.
\newblock
\newblock
\urldef\tempurl%
\url{https://eips.ethereum.org/EIPS/eip-1014}
\showURL{%
Retrieved June 17, 2023 from \tempurl}


\bibitem[Buterin and Reitwiessner(2017)]%
        {eip-197}
\bibfield{author}{\bibinfo{person}{Vitalik Buterin} {and}
  \bibinfo{person}{Christian Reitwiessner}.} \bibinfo{year}{2017}\natexlab{}.
\newblock \bibinfo{title}{EIP-197: Precompiled contracts for optimal ate
  pairing check on the elliptic curve alt\_bn128}.
\newblock
\newblock
\urldef\tempurl%
\url{https://eips.ethereum.org/EIPS/eip-197}
\showURL{%
Retrieved June 17, 2023 from \tempurl}


\bibitem[Caversaccio(2023)]%
        {eip-7266}
\bibfield{author}{\bibinfo{person}{Pascal Caversaccio}.}
  \bibinfo{year}{2023}\natexlab{}.
\newblock \bibinfo{title}{EIP-7266: Remove BLAKE2 compression precompile}.
\newblock
\newblock
\urldef\tempurl%
\url{https://eips.ethereum.org/EIPS/eip-7266}
\showURL{%
Retrieved Oct 17, 2023 from \tempurl}


\bibitem[Chen et~al\mbox{.}(2020)]%
        {chen2020defining}
\bibfield{author}{\bibinfo{person}{Jiachi Chen}, \bibinfo{person}{Xin Xia},
  \bibinfo{person}{David Lo}, \bibinfo{person}{John Grundy},
  \bibinfo{person}{Xiapu Luo}, {and} \bibinfo{person}{Ting Chen}.}
  \bibinfo{year}{2020}\natexlab{}.
\newblock \showarticletitle{Defining smart contract defects on ethereum}.
\newblock \bibinfo{journal}{\emph{IEEE Transactions on Software Engineering}}
  \bibinfo{volume}{48}, \bibinfo{number}{1} (\bibinfo{year}{2020}),
  \bibinfo{pages}{327--345}.
\newblock


\bibitem[Chen et~al\mbox{.}(2021)]%
        {chen2021maintenance}
\bibfield{author}{\bibinfo{person}{Jiachi Chen}, \bibinfo{person}{Xin Xia},
  \bibinfo{person}{David Lo}, \bibinfo{person}{John Grundy}, {and}
  \bibinfo{person}{Xiaohu Yang}.} \bibinfo{year}{2021}\natexlab{}.
\newblock \showarticletitle{Maintenance-related concerns for post-deployed
  Ethereum smart contract development: issues, techniques, and future
  challenges}.
\newblock \bibinfo{journal}{\emph{Empirical Software Engineering}}
  \bibinfo{volume}{26}, \bibinfo{number}{6} (\bibinfo{year}{2021}),
  \bibinfo{pages}{117}.
\newblock


\bibitem[Cohen(1960)]%
        {cohen1960coefficient}
\bibfield{author}{\bibinfo{person}{Jacob Cohen}.}
  \bibinfo{year}{1960}\natexlab{}.
\newblock \showarticletitle{A coefficient of agreement for nominal scales}.
\newblock \bibinfo{journal}{\emph{Educational and psychological measurement}}
  \bibinfo{volume}{20}, \bibinfo{number}{1} (\bibinfo{year}{1960}),
  \bibinfo{pages}{37--46}.
\newblock


\bibitem[community(2023)]%
        {zkroolup}
\bibfield{author}{\bibinfo{person}{Ethereum community}.}
  \bibinfo{year}{2023}\natexlab{}.
\newblock \bibinfo{title}{Zero-knowledge Roolups}.
\newblock
\newblock
\urldef\tempurl%
\url{https://ethereum.org/en/developers/docs/scaling/zk-rollups/}
\showURL{%
Retrieved June 17, 2023 from \tempurl}


\bibitem[Damg{\aa}rd(1998)]%
        {damgaard1998commitment}
\bibfield{author}{\bibinfo{person}{Ivan Damg{\aa}rd}.}
  \bibinfo{year}{1998}\natexlab{}.
\newblock \showarticletitle{Commitment schemes and zero-knowledge protocols}.
  In \bibinfo{booktitle}{\emph{School organized by the European Educational
  Forum}}. Springer, \bibinfo{pages}{63--86}.
\newblock


\bibitem[Dobbertin et~al\mbox{.}(1996)]%
        {dobbertin1996ripemd}
\bibfield{author}{\bibinfo{person}{Hans Dobbertin}, \bibinfo{person}{Antoon
  Bosselaers}, {and} \bibinfo{person}{Bart Preneel}.}
  \bibinfo{year}{1996}\natexlab{}.
\newblock \showarticletitle{RIPEMD-160: A strengthened version of RIPEMD}. In
  \bibinfo{booktitle}{\emph{Fast Software Encryption: Third International
  Workshop Cambridge, UK, February 21--23 1996 Proceedings 3}}. Springer,
  \bibinfo{pages}{71--82}.
\newblock


\bibitem[Eberhardt and Tai(2018)]%
        {eberhardt2018zokrates}
\bibfield{author}{\bibinfo{person}{Jacob Eberhardt} {and}
  \bibinfo{person}{Stefan Tai}.} \bibinfo{year}{2018}\natexlab{}.
\newblock \showarticletitle{Zokrates-scalable privacy-preserving off-chain
  computations}. In \bibinfo{booktitle}{\emph{2018 IEEE International
  Conference on Internet of Things (iThings) and IEEE Green Computing and
  Communications (GreenCom) and IEEE Cyber, Physical and Social Computing
  (CPSCom) and IEEE Smart Data (SmartData)}}. IEEE,
  \bibinfo{pages}{1084--1091}.
\newblock


\bibitem[Egele et~al\mbox{.}(2013)]%
        {egele2013empirical}
\bibfield{author}{\bibinfo{person}{Manuel Egele}, \bibinfo{person}{David
  Brumley}, \bibinfo{person}{Yanick Fratantonio}, {and}
  \bibinfo{person}{Christopher Kruegel}.} \bibinfo{year}{2013}\natexlab{}.
\newblock \showarticletitle{An empirical study of cryptographic misuse in
  android applications}. In \bibinfo{booktitle}{\emph{Proceedings of the 2013
  ACM SIGSAC Conference on Computer \& Communications Security}}.
  \bibinfo{pages}{73--84}.
\newblock


\bibitem[Etherscan(2023)]%
        {etherscan}
\bibfield{author}{\bibinfo{person}{Etherscan}.}
  \bibinfo{year}{2023}\natexlab{}.
\newblock \bibinfo{title}{The Ethereum Blockchain Explorer}.
\newblock
\newblock
\urldef\tempurl%
\url{https://etherscan.io/}
\showURL{%
Retrieved June 17, 2023 from \tempurl}


\bibitem[Exchange(2023a)]%
        {stackexchange}
\bibfield{author}{\bibinfo{person}{Stack Exchange}.}
  \bibinfo{year}{2023}\natexlab{a}.
\newblock \bibinfo{title}{Stack Exchange}.
\newblock
\newblock
\urldef\tempurl%
\url{https://stackexchange.com}
\showURL{%
Retrieved June 17, 2023 from \tempurl}


\bibitem[Exchange(2023b)]%
        {searchapi}
\bibfield{author}{\bibinfo{person}{Stack Exchange}.}
  \bibinfo{year}{2023}\natexlab{b}.
\newblock \bibinfo{title}{Stack Exchange API}.
\newblock
\newblock
\urldef\tempurl%
\url{https://api.stackexchange.com/docs}
\showURL{%
Retrieved June 17, 2023 from \tempurl}


\bibitem[Foundation(2015)]%
        {frontier}
\bibfield{author}{\bibinfo{person}{Ethereum Foundation}.}
  \bibinfo{year}{2015}\natexlab{}.
\newblock \bibinfo{title}{Frontier is coming - what to expect, and how to
  prepare}.
\newblock
\newblock
\urldef\tempurl%
\url{https://blog.ethereum.org/2015/07/22/frontier-is-coming-what-to-expect-and-how-to-prepare}
\showURL{%
Retrieved June 17, 2023 from \tempurl}


\bibitem[Foundation(2017)]%
        {byzantium}
\bibfield{author}{\bibinfo{person}{Ethereum Foundation}.}
  \bibinfo{year}{2017}\natexlab{}.
\newblock \bibinfo{title}{Byzantium HF Announcement}.
\newblock
\newblock
\urldef\tempurl%
\url{https://blog.ethereum.org/2017/10/12/byzantium-hf-announcement}
\showURL{%
Retrieved June 17, 2023 from \tempurl}


\bibitem[Foundation(2019a)]%
        {istanbul}
\bibfield{author}{\bibinfo{person}{Ethereum Foundation}.}
  \bibinfo{year}{2019}\natexlab{a}.
\newblock \bibinfo{title}{Ethereum Istanbul Upgrade Announcement}.
\newblock
\newblock
\urldef\tempurl%
\url{https://blog.ethereum.org/2019/11/20/ethereum-istanbul-upgrade-announcement}
\showURL{%
Retrieved June 17, 2023 from \tempurl}


\bibitem[Foundation(2019b)]%
        {istanbul-upgrade}
\bibfield{author}{\bibinfo{person}{Ethereum Foundation}.}
  \bibinfo{year}{2019}\natexlab{b}.
\newblock \bibinfo{title}{Ethereum Istanbul Upgrade Announcement}.
\newblock
\newblock
\urldef\tempurl%
\url{https://blog.ethereum.org/2019/11/20/ethereum-istanbul-upgrade-announcement}
\showURL{%
Retrieved June 17, 2023 from \tempurl}


\bibitem[Foundation(2023)]%
        {hardfork}
\bibfield{author}{\bibinfo{person}{Ethereum Foundation}.}
  \bibinfo{year}{2023}\natexlab{}.
\newblock \bibinfo{title}{The history of Ethereum}.
\newblock
\newblock
\urldef\tempurl%
\url{https://ethereum.org/en/history}
\showURL{%
Retrieved June 17, 2023 from \tempurl}


\bibitem[Gabizon et~al\mbox{.}(2019)]%
        {gabizon2019plonk}
\bibfield{author}{\bibinfo{person}{Ariel Gabizon}, \bibinfo{person}{Zachary~J
  Williamson}, {and} \bibinfo{person}{Oana Ciobotaru}.}
  \bibinfo{year}{2019}\natexlab{}.
\newblock \showarticletitle{Plonk: Permutations over lagrange-bases for
  oecumenical noninteractive arguments of knowledge}.
\newblock \bibinfo{journal}{\emph{Cryptology ePrint Archive}}
  (\bibinfo{year}{2019}).
\newblock


\bibitem[Gabor et~al\mbox{.}(2007)]%
        {gabor2007types}
\bibfield{author}{\bibinfo{person}{Manuela~Rozalia Gabor} {et~al\mbox{.}}}
  \bibinfo{year}{2007}\natexlab{}.
\newblock \showarticletitle{Types of non-probabilistic sampling used in
  marketing research.„Snowball” sampling}.
\newblock \bibinfo{journal}{\emph{Management \& Marketing-Bucharest}}
  \bibinfo{number}{3} (\bibinfo{year}{2007}), \bibinfo{pages}{80--90}.
\newblock


\bibitem[Github(2019)]%
        {curveSol}
\bibfield{author}{\bibinfo{person}{Github}.} \bibinfo{year}{2019}\natexlab{}.
\newblock \bibinfo{title}{Solidity BN256G2}.
\newblock
\newblock
\urldef\tempurl%
\url{https://github.com/musalbas/solidity-BN256G2}
\showURL{%
Retrieved Oct 17, 2023 from \tempurl}


\bibitem[Github(2023a)]%
        {GithubAPI}
\bibfield{author}{\bibinfo{person}{Github}.} \bibinfo{year}{2023}\natexlab{a}.
\newblock \bibinfo{title}{GitHub REST API documentation}.
\newblock
\newblock
\urldef\tempurl%
\url{https://docs.github.com/en/rest}
\showURL{%
Retrieved June 17, 2023 from \tempurl}


\bibitem[Github(2023b)]%
        {pastaSol}
\bibfield{author}{\bibinfo{person}{Github}.} \bibinfo{year}{2023}\natexlab{b}.
\newblock \bibinfo{title}{Pasta curves in solidity}.
\newblock
\newblock
\urldef\tempurl%
\url{https://github.com/zhenfeizhang/pasta-solidity}
\showURL{%
Retrieved Oct 17, 2023 from \tempurl}


\bibitem[Go-Ethereum(2023)]%
        {debugapi}
\bibfield{author}{\bibinfo{person}{Go-Ethereum}.}
  \bibinfo{year}{2023}\natexlab{}.
\newblock \bibinfo{title}{debug Namespace}.
\newblock
\newblock
\urldef\tempurl%
\url{https://geth.ethereum.org/docs/interacting-with-geth/rpc/ns-debug}
\showURL{%
Retrieved June 17, 2023 from \tempurl}


\bibitem[Green and Smith(2016)]%
        {green2016developers}
\bibfield{author}{\bibinfo{person}{Matthew Green} {and}
  \bibinfo{person}{Matthew Smith}.} \bibinfo{year}{2016}\natexlab{}.
\newblock \showarticletitle{Developers are not the enemy!: The need for usable
  security apis}.
\newblock \bibinfo{journal}{\emph{IEEE Security \& Privacy}}
  \bibinfo{volume}{14}, \bibinfo{number}{5} (\bibinfo{year}{2016}),
  \bibinfo{pages}{40--46}.
\newblock


\bibitem[Groth(2016)]%
        {groth2016size}
\bibfield{author}{\bibinfo{person}{Jens Groth}.}
  \bibinfo{year}{2016}\natexlab{}.
\newblock \showarticletitle{On the size of pairing-based non-interactive
  arguments}. In \bibinfo{booktitle}{\emph{Advances in Cryptology--EUROCRYPT
  2016: 35th Annual International Conference on the Theory and Applications of
  Cryptographic Techniques, Vienna, Austria, May 8-12, 2016, Proceedings, Part
  II 35}}. Springer, \bibinfo{pages}{305--326}.
\newblock


\bibitem[Gu et~al\mbox{.}(2019)]%
        {gu2019empirical}
\bibfield{author}{\bibinfo{person}{Zuxing Gu}, \bibinfo{person}{Jiecheng Wu},
  \bibinfo{person}{Jiaxiang Liu}, \bibinfo{person}{Min Zhou}, {and}
  \bibinfo{person}{Ming Gu}.} \bibinfo{year}{2019}\natexlab{}.
\newblock \showarticletitle{An empirical study on api-misuse bugs in
  open-source c programs}. In \bibinfo{booktitle}{\emph{2019 IEEE 43rd annual
  Computer Software and Applications Conference (COMPSAC)}},
  Vol.~\bibinfo{volume}{1}. IEEE, \bibinfo{pages}{11--20}.
\newblock


\bibitem[Guido et~al\mbox{.}(2011)]%
        {keccak}
\bibfield{author}{\bibinfo{person}{Bertoni Guido}, \bibinfo{person}{Daemen
  Joan}, \bibinfo{person}{Peeters Michal}, {and} \bibinfo{person}{Gilles~Van
  Assche}.} \bibinfo{year}{2011}\natexlab{}.
\newblock \bibinfo{title}{The KECCAK SHA-3 submission}.
\newblock
\newblock
\urldef\tempurl%
\url{https://keccak.team/files/ Keccak-submission-3.pdf}
\showURL{%
Retrieved June 17, 2023 from \tempurl}


\bibitem[Gurkan(2022)]%
        {blssignature}
\bibfield{author}{\bibinfo{person}{Kobi Gurkan}.}
  \bibinfo{year}{2022}\natexlab{}.
\newblock \bibinfo{title}{Optimized BLS multisignatures on EVM}.
\newblock
\newblock
\urldef\tempurl%
\url{https://geometry.xyz/notebook/Optimized-BLS-multisignatures-on-EVM}
\showURL{%
Retrieved June 17, 2023 from \tempurl}


\bibitem[Hazhirpasand et~al\mbox{.}(2020)]%
        {hazhirpasand2020java}
\bibfield{author}{\bibinfo{person}{Mohammadreza Hazhirpasand},
  \bibinfo{person}{Mohammad Ghafari}, {and} \bibinfo{person}{Oscar
  Nierstrasz}.} \bibinfo{year}{2020}\natexlab{}.
\newblock \showarticletitle{Java cryptography uses in the wild}. In
  \bibinfo{booktitle}{\emph{Proceedings of the 14th ACM/IEEE International
  Symposium on Empirical Software Engineering and Measurement (ESEM)}}.
  \bibinfo{pages}{1--6}.
\newblock


\bibitem[Holt et~al\mbox{.}(1980)]%
        {holt1980chi}
\bibfield{author}{\bibinfo{person}{David Holt}, \bibinfo{person}{AJ Scott},
  {and} \bibinfo{person}{PD Ewings}.} \bibinfo{year}{1980}\natexlab{}.
\newblock \showarticletitle{Chi-squared tests with survey data}.
\newblock \bibinfo{journal}{\emph{Journal of the Royal Statistical Society:
  Series A (General)}} \bibinfo{volume}{143}, \bibinfo{number}{3}
  (\bibinfo{year}{1980}), \bibinfo{pages}{303--320}.
\newblock


\bibitem[Hopwood(2020)]%
        {pasta}
\bibfield{author}{\bibinfo{person}{Daira Hopwood}.}
  \bibinfo{year}{2020}\natexlab{}.
\newblock \bibinfo{title}{The Pasta Curves for Halo 2 and Beyond}.
\newblock
\newblock
\urldef\tempurl%
\url{https://electriccoin.co/blog/the-pasta-curves-for-halo-2-and-beyond}
\showURL{%
Retrieved June 17, 2023 from \tempurl}


\bibitem[Hopwood et~al\mbox{.}(2016)]%
        {hopwood2016zcash}
\bibfield{author}{\bibinfo{person}{Daira Hopwood}, \bibinfo{person}{Sean Bowe},
  \bibinfo{person}{Taylor Hornby}, {and} \bibinfo{person}{Nathan Wilcox}.}
  \bibinfo{year}{2016}\natexlab{}.
\newblock \showarticletitle{Zcash protocol specification}.
\newblock \bibinfo{journal}{\emph{GitHub: San Francisco, CA, USA}}
  \bibinfo{volume}{4} (\bibinfo{year}{2016}), \bibinfo{pages}{220}.
\newblock


\bibitem[Ivanov et~al\mbox{.}(2023)]%
        {ivanov2023txt}
\bibfield{author}{\bibinfo{person}{Nikolay Ivanov}, \bibinfo{person}{Qiben
  Yan}, {and} \bibinfo{person}{Anurag Kompalli}.}
  \bibinfo{year}{2023}\natexlab{}.
\newblock \showarticletitle{TxT: Real-time Transaction Encapsulation for
  Ethereum Smart Contracts}.
\newblock \bibinfo{journal}{\emph{IEEE Transactions on Information Forensics
  and Security}}  \bibinfo{volume}{18} (\bibinfo{year}{2023}),
  \bibinfo{pages}{1141--1155}.
\newblock


\bibitem[Johnson et~al\mbox{.}(2001)]%
        {johnson2001elliptic}
\bibfield{author}{\bibinfo{person}{Don Johnson}, \bibinfo{person}{Alfred
  Menezes}, {and} \bibinfo{person}{Scott Vanstone}.}
  \bibinfo{year}{2001}\natexlab{}.
\newblock \showarticletitle{The elliptic curve digital signature algorithm
  (ECDSA)}.
\newblock \bibinfo{journal}{\emph{International Journal of Information
  Security}}  \bibinfo{volume}{1} (\bibinfo{year}{2001}),
  \bibinfo{pages}{36--63}.
\newblock


\bibitem[Joshi et~al\mbox{.}(2015)]%
        {joshi2015likert}
\bibfield{author}{\bibinfo{person}{Ankur Joshi}, \bibinfo{person}{Saket Kale},
  \bibinfo{person}{Satish Chandel}, {and} \bibinfo{person}{D~Kumar Pal}.}
  \bibinfo{year}{2015}\natexlab{}.
\newblock \showarticletitle{Likert scale: Explored and explained}.
\newblock \bibinfo{journal}{\emph{British Journal of Applied Science \&
  Technology}} \bibinfo{volume}{7}, \bibinfo{number}{4} (\bibinfo{year}{2015}),
  \bibinfo{pages}{396}.
\newblock


\bibitem[Kaleem and Shi(2021)]%
        {kaleem2021demystifying}
\bibfield{author}{\bibinfo{person}{Mudabbir Kaleem} {and}
  \bibinfo{person}{Weidong Shi}.} \bibinfo{year}{2021}\natexlab{}.
\newblock \showarticletitle{Demystifying pythia: A survey of chainlink oracles
  usage on ethereum}. In \bibinfo{booktitle}{\emph{Financial Cryptography and
  Data Security. FC 2021 International Workshops: CoDecFin, DeFi, VOTING, and
  WTSC, Virtual Event, March 5, 2021, Revised Selected Papers 25}}. Springer,
  \bibinfo{pages}{115--123}.
\newblock


\bibitem[Kate et~al\mbox{.}(2010)]%
        {kate2010constant}
\bibfield{author}{\bibinfo{person}{Aniket Kate}, \bibinfo{person}{Gregory~M
  Zaverucha}, {and} \bibinfo{person}{Ian Goldberg}.}
  \bibinfo{year}{2010}\natexlab{}.
\newblock \showarticletitle{Constant-size commitments to polynomials and their
  applications}. In \bibinfo{booktitle}{\emph{Advances in Cryptology-ASIACRYPT
  2010: 16th International Conference on the Theory and Application of
  Cryptology and Information Security, Singapore, December 5-9, 2010.
  Proceedings 16}}. Springer, \bibinfo{pages}{177--194}.
\newblock


\bibitem[Kitchenham and Pfleeger(2008)]%
        {kitchenham2008personal}
\bibfield{author}{\bibinfo{person}{Barbara~A Kitchenham} {and}
  \bibinfo{person}{Shari~L Pfleeger}.} \bibinfo{year}{2008}\natexlab{}.
\newblock \showarticletitle{Personal opinion surveys}.
\newblock \bibinfo{journal}{\emph{Guide to advanced empirical software
  engineering}} (\bibinfo{year}{2008}), \bibinfo{pages}{63--92}.
\newblock


\bibitem[Kuszmaul(2019)]%
        {kuszmaul2019verkle}
\bibfield{author}{\bibinfo{person}{John Kuszmaul}.}
  \bibinfo{year}{2019}\natexlab{}.
\newblock \showarticletitle{Verkle trees}.
\newblock  (\bibinfo{year}{2019}).
\newblock


\bibitem[Liao et~al\mbox{.}(2022)]%
        {liao2022large}
\bibfield{author}{\bibinfo{person}{Zhou Liao}, \bibinfo{person}{Shuwei Song},
  \bibinfo{person}{Hang Zhu}, \bibinfo{person}{Xiapu Luo},
  \bibinfo{person}{Zheyuan He}, \bibinfo{person}{Renkai Jiang},
  \bibinfo{person}{Ting Chen}, \bibinfo{person}{Jiachi Chen},
  \bibinfo{person}{Tao Zhang}, {and} \bibinfo{person}{Xiaosong Zhang}.}
  \bibinfo{year}{2022}\natexlab{}.
\newblock \showarticletitle{Large-scale empirical study of inline assembly on
  7.6 million ethereum smart contracts}.
\newblock \bibinfo{journal}{\emph{IEEE Transactions on Software Engineering}}
  \bibinfo{volume}{49}, \bibinfo{number}{2} (\bibinfo{year}{2022}),
  \bibinfo{pages}{777--801}.
\newblock


\bibitem[Liu et~al\mbox{.}(2021)]%
        {liu2021characterizing}
\bibfield{author}{\bibinfo{person}{Lu Liu}, \bibinfo{person}{Lili Wei},
  \bibinfo{person}{Wuqi Zhang}, \bibinfo{person}{Ming Wen},
  \bibinfo{person}{Yepang Liu}, {and} \bibinfo{person}{Shing-Chi Cheung}.}
  \bibinfo{year}{2021}\natexlab{}.
\newblock \showarticletitle{Characterizing transaction-reverting statements in
  ethereum smart contracts}. In \bibinfo{booktitle}{\emph{2021 36th IEEE/ACM
  International Conference on Automated Software Engineering (ASE)}}. IEEE,
  \bibinfo{pages}{630--641}.
\newblock


\bibitem[Lundfal(2020)]%
        {eip-2612}
\bibfield{author}{\bibinfo{person}{Martin Lundfal}.}
  \bibinfo{year}{2020}\natexlab{}.
\newblock \bibinfo{title}{ERC-2612: Permit Extension for EIP-20 Signed
  Approvals}.
\newblock
\newblock
\urldef\tempurl%
\url{https://eips.ethereum.org/EIPS/eip-2612}
\showURL{%
Retrieved June 17, 2023 from \tempurl}


\bibitem[Mindermann et~al\mbox{.}(2018)]%
        {mindermann2018usable}
\bibfield{author}{\bibinfo{person}{Kai Mindermann}, \bibinfo{person}{Philipp
  Keck}, {and} \bibinfo{person}{Stefan Wagner}.}
  \bibinfo{year}{2018}\natexlab{}.
\newblock \showarticletitle{How usable are rust cryptography APIs?}. In
  \bibinfo{booktitle}{\emph{2018 IEEE International Conference on Software
  Quality, Reliability and Security (QRS)}}. IEEE, \bibinfo{pages}{143--154}.
\newblock


\bibitem[Mironov(2006)]%
        {mironov2006collision}
\bibfield{author}{\bibinfo{person}{Ilya Mironov}.}
  \bibinfo{year}{2006}\natexlab{}.
\newblock \showarticletitle{Collision-resistant no more: Hash-and-sign paradigm
  revisited}. In \bibinfo{booktitle}{\emph{Public Key Cryptography-PKC 2006:
  9th International Conference on Theory and Practice in Public-Key
  Cryptography, New York, NY, USA, April 24-26, 2006. Proceedings 9}}.
  Springer, \bibinfo{pages}{140--156}.
\newblock


\bibitem[Nadi et~al\mbox{.}(2016)]%
        {nadi2016jumping}
\bibfield{author}{\bibinfo{person}{Sarah Nadi}, \bibinfo{person}{Stefan
  Kr{\"u}ger}, \bibinfo{person}{Mira Mezini}, {and} \bibinfo{person}{Eric
  Bodden}.} \bibinfo{year}{2016}\natexlab{}.
\newblock \showarticletitle{Jumping through hoops: Why do Java developers
  struggle with cryptography APIs?}. In \bibinfo{booktitle}{\emph{Proceedings
  of the 38th International Conference on Software Engineering}}.
  \bibinfo{pages}{935--946}.
\newblock


\bibitem[Nakamoto(2008)]%
        {nakamoto2008bitcoin}
\bibfield{author}{\bibinfo{person}{Satoshi Nakamoto}.}
  \bibinfo{year}{2008}\natexlab{}.
\newblock \showarticletitle{Bitcoin: A peer-to-peer electronic cash system}.
\newblock \bibinfo{journal}{\emph{Decentralized business review}}
  (\bibinfo{year}{2008}).
\newblock


\bibitem[Nist(1992)]%
        {nist1992digital}
\bibfield{author}{\bibinfo{person}{Corporate Nist}.}
  \bibinfo{year}{1992}\natexlab{}.
\newblock \showarticletitle{The digital signature standard}.
\newblock \bibinfo{journal}{\emph{Commun. ACM}} \bibinfo{volume}{35},
  \bibinfo{number}{7} (\bibinfo{year}{1992}), \bibinfo{pages}{36--40}.
\newblock


\bibitem[of~Standards and (NIST)(2012)]%
        {sha3contest}
\bibfield{author}{\bibinfo{person}{The National~Institute of Standards} {and}
  \bibinfo{person}{Technology (NIST)}.} \bibinfo{year}{2012}\natexlab{}.
\newblock \bibinfo{title}{SHA-3 Selection Announcement}.
\newblock
\newblock
\urldef\tempurl%
\url{https://csrc.nist.gov/CSRC/media/Projects/Hash-Functions/documents/sha-3_selection_announcement.pdf}
\showURL{%
Retrieved June 17, 2023 from \tempurl}


\bibitem[Openzepplin(2023a)]%
        {openzepplinECDSA}
\bibfield{author}{\bibinfo{person}{Openzepplin}.}
  \bibinfo{year}{2023}\natexlab{a}.
\newblock \bibinfo{title}{Checking Signatures On-Chain}.
\newblock
\newblock
\urldef\tempurl%
\url{https://docs.openzeppelin.com/contracts/2.x/utilities}
\showURL{%
Retrieved June 17, 2023 from \tempurl}


\bibitem[Openzepplin(2023b)]%
        {openzepplin}
\bibfield{author}{\bibinfo{person}{Openzepplin}.}
  \bibinfo{year}{2023}\natexlab{b}.
\newblock \bibinfo{title}{The standard for secure blockchain applications}.
\newblock
\newblock
\urldef\tempurl%
\url{https://www.openzeppelin.com/}
\showURL{%
Retrieved June 17, 2023 from \tempurl}


\bibitem[Overflow(2019)]%
        {q56124326q}
\bibfield{author}{\bibinfo{person}{Stack Overflow}.}
  \bibinfo{year}{2019}\natexlab{}.
\newblock \bibinfo{title}{How to decode SHA256 hash value and retrieve data in
  Solidity}.
\newblock
\newblock
\urldef\tempurl%
\url{https://stackoverflow.com/questions/56124326}
\showURL{%
Retrieved June 17, 2023 from \tempurl}


\bibitem[Overflow(2023)]%
        {stackoverflow}
\bibfield{author}{\bibinfo{person}{Stack Overflow}.}
  \bibinfo{year}{2023}\natexlab{}.
\newblock \bibinfo{title}{Stack Overflow}.
\newblock
\newblock
\urldef\tempurl%
\url{https://stackoverflow.com}
\showURL{%
Retrieved June 17, 2023 from \tempurl}


\bibitem[Ozcelik et~al\mbox{.}(2021)]%
        {ozcelik2021overview}
\bibfield{author}{\bibinfo{person}{Ilker Ozcelik}, \bibinfo{person}{Sai
  Medury}, \bibinfo{person}{Justin Broaddus}, {and} \bibinfo{person}{Anthony
  Skjellum}.} \bibinfo{year}{2021}\natexlab{}.
\newblock \showarticletitle{An Overview of Cryptographic Accumulators}.
\newblock \bibinfo{journal}{\emph{arXiv preprint arXiv:2103.04330}}
  (\bibinfo{year}{2021}).
\newblock


\bibitem[Patnaik et~al\mbox{.}(2019)]%
        {patnaik2019usability}
\bibfield{author}{\bibinfo{person}{Nikhil Patnaik}, \bibinfo{person}{Joseph
  Hallett}, {and} \bibinfo{person}{Awais Rashid}.}
  \bibinfo{year}{2019}\natexlab{}.
\newblock \showarticletitle{Usability Smells: An Analysis of Developers'
  Struggle With Crypto Libraries.}. In \bibinfo{booktitle}{\emph{SOUPS@ USENIX
  Security Symposium}}.
\newblock


\bibitem[Penard and van Werkhoven(2008)]%
        {penard2008secure}
\bibfield{author}{\bibinfo{person}{Wouter Penard} {and} \bibinfo{person}{Tim
  van Werkhoven}.} \bibinfo{year}{2008}\natexlab{}.
\newblock \showarticletitle{On the secure hash algorithm family}.
\newblock \bibinfo{journal}{\emph{Cryptography in context}}
  (\bibinfo{year}{2008}), \bibinfo{pages}{1--18}.
\newblock


\bibitem[Piantadosi et~al\mbox{.}(2023)]%
        {piantadosi2023detecting}
\bibfield{author}{\bibinfo{person}{Valentina Piantadosi},
  \bibinfo{person}{Giovanni Rosa}, \bibinfo{person}{Davide Placella},
  \bibinfo{person}{Simone Scalabrino}, {and} \bibinfo{person}{Rocco Oliveto}.}
  \bibinfo{year}{2023}\natexlab{}.
\newblock \showarticletitle{Detecting functional and security-related issues in
  smart contracts: A systematic literature review}.
\newblock \bibinfo{journal}{\emph{Software: Practice and Experience}}
  \bibinfo{volume}{53}, \bibinfo{number}{2} (\bibinfo{year}{2023}),
  \bibinfo{pages}{465--495}.
\newblock


\bibitem[Preneel(1994)]%
        {preneel1994cryptographic}
\bibfield{author}{\bibinfo{person}{Bart Preneel}.}
  \bibinfo{year}{1994}\natexlab{}.
\newblock \showarticletitle{Cryptographic hash functions}.
\newblock \bibinfo{journal}{\emph{European Transactions on Telecommunications}}
  \bibinfo{volume}{5}, \bibinfo{number}{4} (\bibinfo{year}{1994}),
  \bibinfo{pages}{431--448}.
\newblock


\bibitem[Registry(2023a)]%
        {swclist}
\bibfield{author}{\bibinfo{person}{SWC Registry}.}
  \bibinfo{year}{2023}\natexlab{a}.
\newblock \bibinfo{title}{Smart Contract Weakness Classification and Test
  Cases}.
\newblock
\newblock
\urldef\tempurl%
\url{https://swcregistry.io/}
\showURL{%
Retrieved June 17, 2023 from \tempurl}


\bibitem[Registry(2023b)]%
        {swc-120}
\bibfield{author}{\bibinfo{person}{SWC Registry}.}
  \bibinfo{year}{2023}\natexlab{b}.
\newblock \bibinfo{title}{Weak Sources of Randomness from Chain Attributes}.
\newblock
\newblock
\urldef\tempurl%
\url{https://swcregistry.io/docs/SWC-120}
\showURL{%
Retrieved June 17, 2023 from \tempurl}


\bibitem[Reitwiessner(2017)]%
        {eip-196}
\bibfield{author}{\bibinfo{person}{Christian Reitwiessner}.}
  \bibinfo{year}{2017}\natexlab{}.
\newblock \bibinfo{title}{EIP-196: Precompiled contracts for addition and
  scalar multiplication on the elliptic curve alt\_bn128}.
\newblock
\newblock
\urldef\tempurl%
\url{https://eips.ethereum.org/EIPS/eip-196}
\showURL{%
Retrieved June 17, 2023 from \tempurl}


\bibitem[Rivest et~al\mbox{.}(1978)]%
        {rivest1978method}
\bibfield{author}{\bibinfo{person}{Ronald~L Rivest}, \bibinfo{person}{Adi
  Shamir}, {and} \bibinfo{person}{Leonard Adleman}.}
  \bibinfo{year}{1978}\natexlab{}.
\newblock \showarticletitle{A method for obtaining digital signatures and
  public-key cryptosystems}.
\newblock \bibinfo{journal}{\emph{Commun. ACM}} \bibinfo{volume}{21},
  \bibinfo{number}{2} (\bibinfo{year}{1978}), \bibinfo{pages}{120--126}.
\newblock


\bibitem[Robillard(2009)]%
        {robillard2009makes}
\bibfield{author}{\bibinfo{person}{Martin~P Robillard}.}
  \bibinfo{year}{2009}\natexlab{}.
\newblock \showarticletitle{What makes APIs hard to learn? Answers from
  developers}.
\newblock \bibinfo{journal}{\emph{IEEE software}} \bibinfo{volume}{26},
  \bibinfo{number}{6} (\bibinfo{year}{2009}), \bibinfo{pages}{27--34}.
\newblock


\bibitem[Schnorr(1990)]%
        {schnorr1990efficient}
\bibfield{author}{\bibinfo{person}{Claus-Peter Schnorr}.}
  \bibinfo{year}{1990}\natexlab{}.
\newblock \showarticletitle{Efficient identification and signatures for smart
  cards}. In \bibinfo{booktitle}{\emph{Advances in Cryptology—CRYPTO’89
  Proceedings 9}}. Springer, \bibinfo{pages}{239--252}.
\newblock


\bibitem[Sendner et~al\mbox{.}(2023)]%
        {sendner2023smarter}
\bibfield{author}{\bibinfo{person}{Christoph Sendner}, \bibinfo{person}{Huili
  Chen}, \bibinfo{person}{Hossein Fereidooni}, \bibinfo{person}{Lukas Petzi},
  \bibinfo{person}{Jan K{\"o}nig}, \bibinfo{person}{Jasper Stang},
  \bibinfo{person}{Alexandra Dmitrienko}, \bibinfo{person}{Ahmad-Reza Sadeghi},
  {and} \bibinfo{person}{Farinaz Koushanfar}.} \bibinfo{year}{2023}\natexlab{}.
\newblock \showarticletitle{Smarter Contracts: Detecting Vulnerabilities in
  Smart Contracts with Deep Transfer Learning.}. In
  \bibinfo{booktitle}{\emph{NDSS}}.
\newblock


\bibitem[Song et~al\mbox{.}(2022)]%
        {song2022esbmc}
\bibfield{author}{\bibinfo{person}{Kunjian Song}, \bibinfo{person}{Nedas
  Matulevicius}, \bibinfo{person}{Eddie~B de Lima~Filho}, {and}
  \bibinfo{person}{Lucas~C Cordeiro}.} \bibinfo{year}{2022}\natexlab{}.
\newblock \showarticletitle{ESBMC-solidity: an SMT-based model checker for
  solidity smart contracts}. In \bibinfo{booktitle}{\emph{Proceedings of the
  ACM/IEEE 44th International Conference on Software Engineering: Companion
  Proceedings}}. \bibinfo{pages}{65--69}.
\newblock


\bibitem[StackExchange(2018)]%
        {49261}
\bibfield{author}{\bibinfo{person}{Ethereum StackExchange}.}
  \bibinfo{year}{2018}\natexlab{}.
\newblock \bibinfo{title}{How to use ecrecover() and what it is?}
\newblock
\newblock
\urldef\tempurl%
\url{https://ethereum.stackexchange.com/questions/49261}
\showURL{%
Retrieved June 17, 2023 from \tempurl}


\bibitem[StackExchange(2019)]%
        {73744}
\bibfield{author}{\bibinfo{person}{Ethereum StackExchange}.}
  \bibinfo{year}{2019}\natexlab{}.
\newblock \bibinfo{title}{How to do RSA signature verification based on
  eip-198}.
\newblock
\newblock
\urldef\tempurl%
\url{https://ethereum.stackexchange.com/questions/73744}
\showURL{%
Retrieved June 17, 2023 from \tempurl}


\bibitem[StackExchange(2021)]%
        {112807}
\bibfield{author}{\bibinfo{person}{Ethereum StackExchange}.}
  \bibinfo{year}{2021}\natexlab{}.
\newblock \bibinfo{title}{Using ECDSA.sol to Sign smartcontract}.
\newblock
\newblock
\urldef\tempurl%
\url{https://ethereum.stackexchange.com/questions/112807}
\showURL{%
Retrieved June 17, 2023 from \tempurl}


\bibitem[StackExchange(2023)]%
        {ethereumstackexchange}
\bibfield{author}{\bibinfo{person}{Ethereum StackExchange}.}
  \bibinfo{year}{2023}\natexlab{}.
\newblock \bibinfo{title}{Ethereum StackExchange}.
\newblock
\newblock
\urldef\tempurl%
\url{https://ethereum.stackexchange.com}
\showURL{%
Retrieved June 17, 2023 from \tempurl}


\bibitem[Steffen et~al\mbox{.}(2022)]%
        {steffen2022zeestar}
\bibfield{author}{\bibinfo{person}{Samuel Steffen}, \bibinfo{person}{Benjamin
  Bichsel}, \bibinfo{person}{Roger Baumgartner}, {and} \bibinfo{person}{Martin
  Vechev}.} \bibinfo{year}{2022}\natexlab{}.
\newblock \showarticletitle{Zeestar: Private smart contracts by homomorphic
  encryption and zero-knowledge proofs}. In \bibinfo{booktitle}{\emph{2022 IEEE
  Symposium on Security and Privacy (SP)}}. IEEE, \bibinfo{pages}{179--197}.
\newblock


\bibitem[Steffen et~al\mbox{.}(2019)]%
        {steffen2019zkay}
\bibfield{author}{\bibinfo{person}{Samuel Steffen}, \bibinfo{person}{Benjamin
  Bichsel}, \bibinfo{person}{Mario Gersbach}, \bibinfo{person}{Noa Melchior},
  \bibinfo{person}{Petar Tsankov}, {and} \bibinfo{person}{Martin Vechev}.}
  \bibinfo{year}{2019}\natexlab{}.
\newblock \showarticletitle{zkay: Specifying and enforcing data privacy in
  smart contracts}. In \bibinfo{booktitle}{\emph{Proceedings of the 2019 ACM
  SIGSAC Conference on Computer and Communications Security}}.
  \bibinfo{pages}{1759--1776}.
\newblock


\bibitem[Team(2023)]%
        {goethereum}
\bibfield{author}{\bibinfo{person}{Go~Ethereum Team}.}
  \bibinfo{year}{2023}\natexlab{}.
\newblock \bibinfo{title}{Go Ethereum}.
\newblock
\newblock
\urldef\tempurl%
\url{https://github.com/ethereum/go-ethereum/tree/master}
\showURL{%
Retrieved Oct 17, 2023 from \tempurl}


\bibitem[Thaler et~al\mbox{.}(2022)]%
        {thaler2022proofs}
\bibfield{author}{\bibinfo{person}{Justin Thaler} {et~al\mbox{.}}}
  \bibinfo{year}{2022}\natexlab{}.
\newblock \showarticletitle{Proofs, arguments, and zero-knowledge}.
\newblock \bibinfo{journal}{\emph{Foundations and Trends{\textregistered} in
  Privacy and Security}} \bibinfo{volume}{4}, \bibinfo{number}{2--4}
  (\bibinfo{year}{2022}), \bibinfo{pages}{117--660}.
\newblock


\bibitem[Tjaden et~al\mbox{.}(2016)]%
        {eip-152}
\bibfield{author}{\bibinfo{person}{Hess Tjaden}, \bibinfo{person}{Luongo Matt},
  \bibinfo{person}{Dyraga Piotr}, {and} \bibinfo{person}{Hancock James}.}
  \bibinfo{year}{2016}\natexlab{}.
\newblock \bibinfo{title}{EIP-152: Add BLAKE2 compression function `F`
  precompile}.
\newblock
\newblock
\urldef\tempurl%
\url{https://eips.ethereum.org/EIPS/eip-152}
\showURL{%
Retrieved June 17, 2023 from \tempurl}


\bibitem[Tyagi(1989)]%
        {tyagi1989effects}
\bibfield{author}{\bibinfo{person}{Pradeep~K Tyagi}.}
  \bibinfo{year}{1989}\natexlab{}.
\newblock \showarticletitle{The effects of appeals, anonymity, and feedback on
  mail survey response patterns from salespeople}.
\newblock \bibinfo{journal}{\emph{Journal of the Academy of Marketing Science}}
   \bibinfo{volume}{17} (\bibinfo{year}{1989}), \bibinfo{pages}{235--241}.
\newblock


\bibitem[Vitalik et~al\mbox{.}(2022)]%
        {eip-4844}
\bibfield{author}{\bibinfo{person}{Buterin Vitalik}, \bibinfo{person}{Feist
  Dankrad}, \bibinfo{person}{Loerakker Diederik}, \bibinfo{person}{Kadianakis
  George}, \bibinfo{person}{Garnett Matt}, \bibinfo{person}{Taiwo Mofi}, {and}
  \bibinfo{person}{Dietrichs Ansgar}.} \bibinfo{year}{2022}\natexlab{}.
\newblock \bibinfo{title}{EIP-4844: Shard Blob Transactions}.
\newblock
\newblock
\urldef\tempurl%
\url{https://eips.ethereum.org/EIPS/eip-4844}
\showURL{%
Retrieved June 17, 2023 from \tempurl}


\bibitem[Wan et~al\mbox{.}(2021)]%
        {wan2021smart}
\bibfield{author}{\bibinfo{person}{Zhiyuan Wan}, \bibinfo{person}{Xin Xia},
  \bibinfo{person}{David Lo}, \bibinfo{person}{Jiachi Chen},
  \bibinfo{person}{Xiapu Luo}, {and} \bibinfo{person}{Xiaohu Yang}.}
  \bibinfo{year}{2021}\natexlab{}.
\newblock \showarticletitle{Smart contract security: a practitioners'
  perspective}. In \bibinfo{booktitle}{\emph{2021 IEEE/ACM 43rd International
  Conference on Software Engineering (ICSE)}}. IEEE,
  \bibinfo{pages}{1410--1422}.
\newblock


\bibitem[Wang et~al\mbox{.}(2023)]%
        {wang2023zero}
\bibfield{author}{\bibinfo{person}{Zhipeng Wang}, \bibinfo{person}{Stefanos
  Chaliasos}, \bibinfo{person}{Kaihua Qin}, \bibinfo{person}{Liyi Zhou},
  \bibinfo{person}{Lifeng Gao}, \bibinfo{person}{Pascal Berrang},
  \bibinfo{person}{Benjamin Livshits}, {and} \bibinfo{person}{Arthur Gervais}.}
  \bibinfo{year}{2023}\natexlab{}.
\newblock \showarticletitle{On how zero-knowledge proof blockchain mixers
  improve, and worsen user privacy}. In \bibinfo{booktitle}{\emph{Proceedings
  of the ACM Web Conference 2023}}. \bibinfo{pages}{2022--2032}.
\newblock


\bibitem[Weiss(2004)]%
        {weiss2004java}
\bibfield{author}{\bibinfo{person}{Jason Weiss}.}
  \bibinfo{year}{2004}\natexlab{}.
\newblock \bibinfo{booktitle}{\emph{Java Cryptography Extensions: Practical
  Guide for Programmers}}.
\newblock \bibinfo{publisher}{Morgan Kaufmann Publishers Inc.},
  \bibinfo{address}{San Francisco, CA, USA}.
\newblock
\showISBNx{0127427511}


\bibitem[Wiki(2019)]%
        {secp256k1}
\bibfield{author}{\bibinfo{person}{Bitcoin Wiki}.}
  \bibinfo{year}{2019}\natexlab{}.
\newblock \bibinfo{title}{Secp256k1}.
\newblock
\newblock
\urldef\tempurl%
\url{https://en.bitcoin.it/wiki/Secp256k1}
\showURL{%
Retrieved June 17, 2023 from \tempurl}


\bibitem[Wikipedia(2023a)]%
        {random-number}
\bibfield{author}{\bibinfo{person}{Wikipedia}.}
  \bibinfo{year}{2023}\natexlab{a}.
\newblock \bibinfo{title}{List of random number generators}.
\newblock
\newblock
\urldef\tempurl%
\url{https://en.wikipedia.org/wiki/List_of_random_number_generators}
\showURL{%
Retrieved June 17, 2023 from \tempurl}


\bibitem[Wikipedia(2023b)]%
        {proof-of-work}
\bibfield{author}{\bibinfo{person}{Wikipedia}.}
  \bibinfo{year}{2023}\natexlab{b}.
\newblock \bibinfo{title}{Proof of work}.
\newblock
\newblock
\urldef\tempurl%
\url{https://en.wikipedia.org/wiki/Proof_of_work}
\showURL{%
Retrieved June 17, 2023 from \tempurl}


\bibitem[Wood et~al\mbox{.}(2014)]%
        {wood2014ethereum}
\bibfield{author}{\bibinfo{person}{Gavin Wood} {et~al\mbox{.}}}
  \bibinfo{year}{2014}\natexlab{}.
\newblock \showarticletitle{Ethereum: A secure decentralised generalised
  transaction ledger}.
\newblock \bibinfo{journal}{\emph{Ethereum project yellow paper}}
  \bibinfo{volume}{151}, \bibinfo{number}{2014} (\bibinfo{year}{2014}),
  \bibinfo{pages}{1--32}.
\newblock


\bibitem[Wood and Wood(2008)]%
        {wood2008card}
\bibfield{author}{\bibinfo{person}{Jed~R Wood} {and} \bibinfo{person}{Larry~E
  Wood}.} \bibinfo{year}{2008}\natexlab{}.
\newblock \showarticletitle{Card sorting: current practices and beyond}.
\newblock \bibinfo{journal}{\emph{Journal of Usability Studies}}
  \bibinfo{volume}{4}, \bibinfo{number}{1} (\bibinfo{year}{2008}),
  \bibinfo{pages}{1--6}.
\newblock


\bibitem[Yang et~al\mbox{.}(2023)]%
        {yang2023definition}
\bibfield{author}{\bibinfo{person}{Shuo Yang}, \bibinfo{person}{Jiachi Chen},
  {and} \bibinfo{person}{Zibin Zheng}.} \bibinfo{year}{2023}\natexlab{}.
\newblock \showarticletitle{Definition and Detection of Defects in NFT Smart
  Contracts}. In \bibinfo{booktitle}{\emph{32nd ACM SIGSOFT International
  Symposium on Software Testing and Analysis}}.
\newblock


\bibitem[Zhang et~al\mbox{.}(2021)]%
        {zhang2021boros}
\bibfield{author}{\bibinfo{person}{Jingjing Zhang}, \bibinfo{person}{Yongjie
  Ye}, \bibinfo{person}{Weigang Wu}, {and} \bibinfo{person}{Xiapu Luo}.}
  \bibinfo{year}{2021}\natexlab{}.
\newblock \showarticletitle{Boros: Secure and efficient off-blockchain
  transactions via payment channel hub}.
\newblock \bibinfo{journal}{\emph{IEEE Transactions on Dependable and Secure
  Computing}} (\bibinfo{year}{2021}).
\newblock


\bibitem[Zhang et~al\mbox{.}(2022)]%
        {zhang2022automatic}
\bibfield{author}{\bibinfo{person}{Ying Zhang}, \bibinfo{person}{Md~Mahir~Asef
  Kabir}, \bibinfo{person}{Ya Xiao}, \bibinfo{person}{Danfeng Yao}, {and}
  \bibinfo{person}{Na Meng}.} \bibinfo{year}{2022}\natexlab{}.
\newblock \showarticletitle{Automatic Detection of Java Cryptographic API
  Misuses: Are We There Yet?}
\newblock \bibinfo{journal}{\emph{IEEE Transactions on Software Engineering}}
  \bibinfo{volume}{49}, \bibinfo{number}{1} (\bibinfo{year}{2022}),
  \bibinfo{pages}{288--303}.
\newblock


\bibitem[Zhang et~al\mbox{.}(2023)]%
        {zhang2023demystifying}
\bibfield{author}{\bibinfo{person}{Zhuo Zhang}, \bibinfo{person}{Brian Zhang},
  \bibinfo{person}{Wen Xu}, {and} \bibinfo{person}{Zhiqiang Lin}.}
  \bibinfo{year}{2023}\natexlab{}.
\newblock \showarticletitle{Demystifying Exploitable Bugs in Smart Contracts}.
  In \bibinfo{booktitle}{\emph{45th International Conference on Software
  Engineering (ICSE)}}.
\newblock


\bibitem[Zou et~al\mbox{.}(2019)]%
        {zou2019smart}
\bibfield{author}{\bibinfo{person}{Weiqin Zou}, \bibinfo{person}{David Lo},
  \bibinfo{person}{Pavneet~Singh Kochhar}, \bibinfo{person}{Xuan-Bach~Dinh Le},
  \bibinfo{person}{Xin Xia}, \bibinfo{person}{Yang Feng},
  \bibinfo{person}{Zhenyu Chen}, {and} \bibinfo{person}{Baowen Xu}.}
  \bibinfo{year}{2019}\natexlab{}.
\newblock \showarticletitle{Smart contract development: Challenges and
  opportunities}.
\newblock \bibinfo{journal}{\emph{IEEE Transactions on Software Engineering}}
  \bibinfo{volume}{47}, \bibinfo{number}{10} (\bibinfo{year}{2019}),
  \bibinfo{pages}{2084--2106}.
\newblock


\end{thebibliography}

\appendix

\end{document}